\documentclass[manuscript=article]{achemso}

\setkeys{acs}{usetitle=true}

\usepackage{caption,subcaption,amsmath,xspace,booktabs,siunitx,multirow,pdflscape,longtable}
\usepackage[dvipsnames]{xcolor}
\usepackage[version=3]{mhchem}

\usepackage{xr}

\makeatletter
\newcommand*{\addFileDependency}[1]{
  \typeout{(#1)}
  \@addtofilelist{#1}
  \IfFileExists{#1}{}{\typeout{No file #1.}}
}
\makeatother

\newcommand*{\extdoc}[1]{%
  \externaldocument{#1}
  \addFileDependency{#1.tex}%
  \addFileDependency{#1.aux}%
}
\extdoc{./suppinfo}

\newcommand{\rrscan}{r$^2$SCAN\xspace}
\newcommand{\angstrom}{\mbox{\normalfont\AA}}
\newcommand{\eform}{\ensuremath{E_\mathrm{f}^0}\xspace}
\newcommand{\ecoh}{\ensuremath{E_\mathrm{coh}}\xspace}
\newcommand{\absforce}{\ensuremath{|\mathbf{F}_i|}\xspace}

\title[MatPES]{A Foundational Potential Energy Surface Dataset for Materials}
\author{Aaron D. Kaplan}
\affiliation[MSD]{Materials Sciences Division, Lawrence Berkeley National Laboratory, Berkeley, California 94720, United States}
\altaffiliation{These authors contributed equally to this work.}

\author{Runze Liu}
\affiliation[UCSD]{Aiiso Yufeng Li Family Department of Chemical and Nano Engineering, University of California San Diego, 9500 Gilman Dr, Mail Code 0448, La Jolla, CA 92093-0448, United States}
\altaffiliation{These authors contributed equally to this work.}

\author{Ji Qi}
\affiliation[UCSD]{Aiiso Yufeng Li Family Department of Chemical and Nano Engineering, University of California San Diego, 9500 Gilman Dr, Mail Code 0448, La Jolla, CA 92093-0448, United States}
\altaffiliation{These authors contributed equally to this work.}

\author{Tsz Wai Ko}
\affiliation[UCSD]{Aiiso Yufeng Li Family Department of Chemical and Nano Engineering, University of California San Diego, 9500 Gilman Dr, Mail Code 0448, La Jolla, CA 92093-0448, United States}

\author{Bowen Deng} 
\affiliation[UCB]{Department of Materials Science and Engineering, University of California, Berkeley, California 94720, United States}
\alsoaffiliation[MSD]{Materials Sciences Division, Lawrence Berkeley National Laboratory, Berkeley, California 94720, United States}

\author{Janosh Riebesell}
\affiliation[MSD]{Materials Sciences Division, Lawrence Berkeley National Laboratory, Berkeley, California 94720, United States}
\alsoaffiliation[CAM]{Cavendish Laboratory, University of Cambridge, J. J. Thomson Ave, Cambridge, UK}

\author{Gerbrand Ceder}
\affiliation[MSD]{Materials Sciences Division, Lawrence Berkeley National Laboratory, Berkeley, California 94720, United States}

\author{Kristin A. Persson}
\affiliation[UCB]{Department of Materials Science and Engineering, University of California, Berkeley, CA 94720, USA}
\alsoaffiliation[MSD]{Materials Sciences Division, Lawrence Berkeley National Laboratory, Berkeley, California 94720, United States}

\author{Shyue Ping Ong}
\email{ongsp@ucsd.edu}
\affiliation[UCSD]{Aiiso Yufeng Li Family Department of Chemical and Nano Engineering, University of California San Diego, 9500 Gilman Dr, Mail Code 0448, La Jolla, CA 92093-0448, United States}
\date{\today}

\begin{document}

\maketitle

\begin{abstract}
Accurate potential energy surface (PES) descriptions are essential for atomistic simulations of materials. Universal machine learning interatomic potentials (UMLIPs)\cite{chen2022m3gnet,deng2023chgnet,batatia2024mace} offer a computationally efficient alternative to density functional theory (DFT)\cite{kohn1965kseq} for PES modeling across the periodic table. However, their accuracy today is fundamentally constrained due to a reliance on DFT relaxation data.\cite{jain2013mp,schmidt2023alex} Here, we introduce MatPES, a foundational PES dataset comprising $\sim 400,000$ structures carefully sampled from 281 million molecular dynamics snapshots that span 16 billion atomic environments. We demonstrate that UMLIPs trained on the modestly sized MatPES dataset can rival, or even outperform, prior models trained on much larger datasets across a broad range of equilibrium, near-equilibrium, and molecular dynamics property benchmarks. We also introduce the first high-fidelity PES dataset based on the revised regularized strongly constrained and appropriately normed (\rrscan) functional \cite{furness2020r2scan} with greatly improved descriptions of interatomic bonding. The open source MatPES initiative emphasizes the importance of data quality over quantity in materials science and enables broad community-driven advancements toward more reliable, generalizable, and efficient UMLIPs for large-scale materials discovery and design.
\end{abstract}

Electronic structure methods, such as those based on Kohn-Sham DFT\cite{kohn1965kseq}, provide the most accurate descriptions of the PES. However, DFT typically scales with the number of electrons cubed, making it prohibitively expensive for simulating complex materials requiring models with large numbers of atoms (e.g., low-symmetry interfaces, amorphous materials, etc.) or properties requiring long time-scale statistics (e.g. diffusivity). To overcome this limitation, an interatomic potential (IP), also known as a force field, is often used to approximate the PES with linear scaling with respect to the number of atoms. Classical IPs, where a functional form is prescribed \cite{lennardjones1931,daw1984eam}, sacrifice substantial accuracy and are limited to a particular chemical identity and bond regime\cite{senftle2016reaxff}.

Machine learning IPs (MLIPs) have emerged as a computationally efficient way to bridge the gap between DFT and classical IPs by using an ML model to learn the DFT PES for different configurations of atoms.\cite{ko2023recent, zhang2024roadmap}
By explicit construction, message passing, or both, MLIPs can capture multi-body interactions to simulate diverse bonding. Among MLIP architectures, graph-based architectures\cite{chen2022m3gnet,batatia2022mace,deng2023chgnet,simeon2023tensornet} have a distinct advantage in handling systems of high compositional complexity by using a unique learned embedding vector\cite{chen2022m3gnet} to represent each unique element. State-of-the-art architectures typically combine message-passing graphs with many-body interactions \cite{qiao2020orbnet,chen2022m3gnet,deng2023chgnet} to achieve an optimal balance between flexibility and efficiency. 

In the past two years, a special class of \textit{universal} MLIPs (UMLIPs) \cite{chen2022m3gnet,deng2023chgnet,batatia2024mace} have emerged with nearly complete coverage of the periodic table. UMLIPs can potentially serve as a drop-in replacement for expensive DFT calculations in a wide range of applications, such as structural relaxations, molecular dynamics (MD) simulations, prediction of PES-derived properties such as phonon dispersions, elastic constants, etc. 

However, present UMLIPs are still limited in their accuracy, especially compared to custom-fitted MLIPs. This can be attributed to three major limitations in often employed data sets. For example, the Materials Project\cite{jain2013mp} structural relaxation dataset (MPF \cite{chen2022m3gnet} or MPtrj \cite{deng2023chgnet}, here referred to collectively as ``MPRelax'') is the most commonly used dataset to train UMLIPs,\cite{riebesell2024mbd} but the rationale for the dataset's creation over the past decade did not prioritize PES accuracy. First, the MPRelax dataset comprises mostly near-equilibrium structures and therefore can only inform the shape of the PES directly adjacent to a minimum. Second, the MPRelax dataset mixes calculations using the Perdew-Burke-Ernzerhof (PBE)\cite{perdew1996pbe} generalized gradient approximation (GGA) exchange-correlation functional without and with a Hubbard $U$ (PBE$+U$) parameter, which are then empirically adjusted to reproduce experimental formation energies.\cite{jain2011mixing} The forces and stresses computed using PBE and PBE+$U$ also differ \cite{shishkin2019dftuforce}, but remain unadjusted, resulting in a mismatch between the treatment of PES quantities. 
This opens the possibility for non-smooth features in the PES when moving between distinct chemical spaces trained on PBE and PBE$+U$ data.
Finally, the computational settings used in MPRelax and other datasets\cite{schmidt2023alex} were chosen to balance computational cost and accuracy in high-throughput structural relaxation workflows, with changes reflecting improvements in DFT methodology made over the course of more than a decade.

The end result is that the MPRelax data contains significant systematic and unsystematic noise in its description of PESs. \citet{qiRobustTrainingMachine2024} demonstrated that the substitution of noisy PES data with accurate single-point DFT calculations can improve the accuracy of UMLIPs, as well as their reliability in molecular dynamics (MD) simulations.
\citet{deng2024soft} have also found that current UMLIPs tend to underpredict larger-magnitude interatomic forces and over-soften phonons, which is likely due to under-sampling of off-equilibrium local environments (i.e., those farther from the PES minimum).

There have been efforts to go beyond the limitations of the MPRelax dataset through brute-force data generation, most notably by industry research groups.\cite{merchant2023gnome,yang2024mattersim,barrosoluque2024omat} For instance, Meta recently released the Open Materials 2024 (OMat24) dataset\cite{barrosoluque2024omat}, which comprises around 100 million structures. However, with the notable exception of OMat24, industry datasets are usually closed source\cite{merchant2023gnome,yang2024mattersim} and inaccessible to the wider research community. Training with such immense datasets also requires resources beyond those readily available at public computing centers.

\begin{figure}[htp]
\centering
\includegraphics[width=\columnwidth]{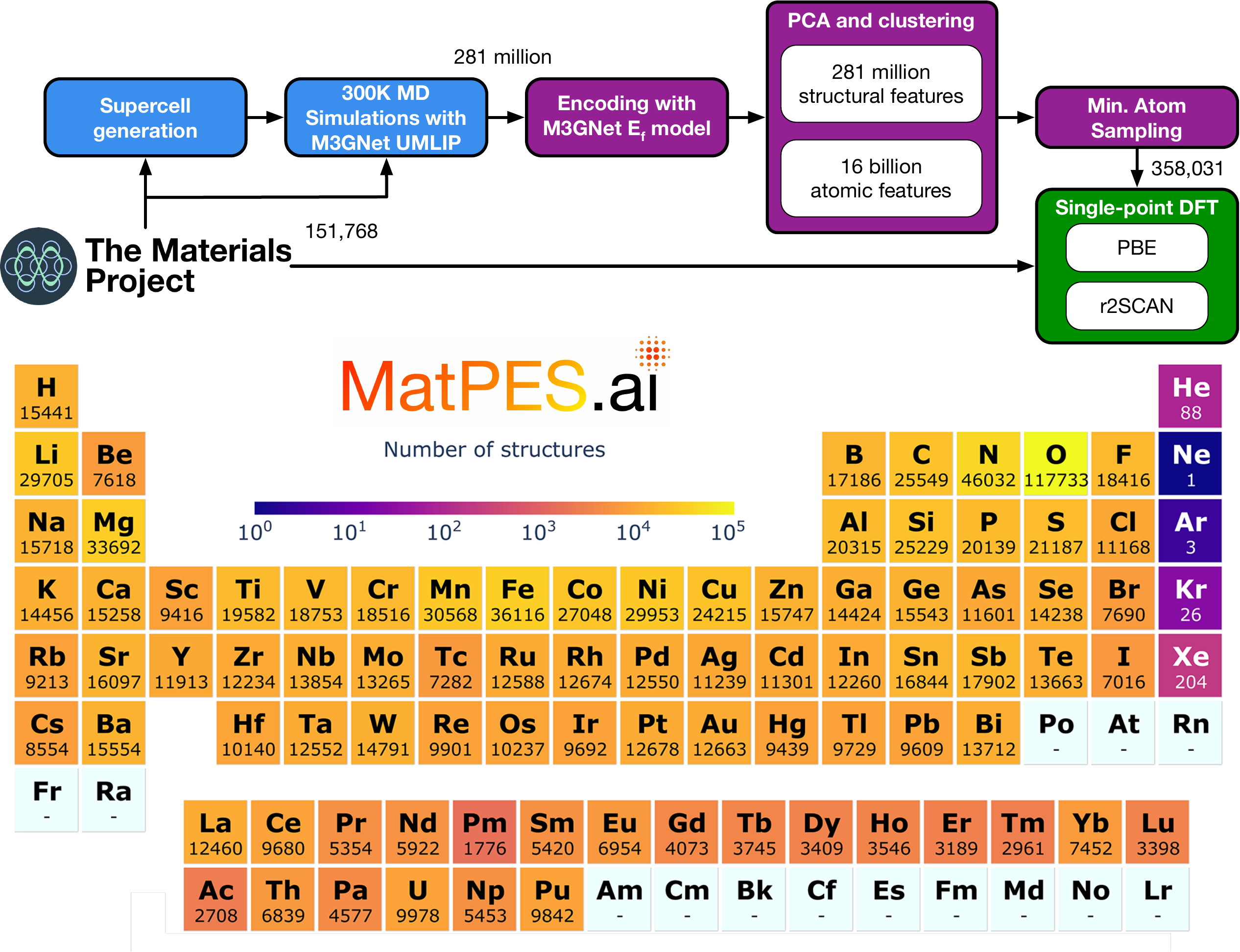}
\caption{
    \label{fig:workflow}\textbf{MatPES dataset development workflow.} The number of structures at each stage in the workflow is indicated. A comprehensive configuration space was generated by performing \textit{NpT} MD simulations at 300K and 1 atm on 281,572 ground-state structures and supercells obtained from the Materials Project (v2022.10.28)\cite{jain2013mp} using a pre-trained M3GNet UMLIP (version MP-2021.2.8-DIRECT). A 2-stage DImensionality-Reduced Encoded Clusters with sTratified (2DIRECT) sampling\cite{qiRobustTrainingMachine2024} was then used to extract representative structures from a configuration space of $\sim$ 281 million structures with $\sim$ 16 billion atomic environments. In each cluster, the structure with the smallest number of atoms was selected to minimize the computational burden. The MD dataset was then augmented with ground-state structures with $< 100$ atoms per cell from the Materials Project to ensure coverage of equilibrium local environments. Single-point DFT calculations with stringent energy and force convergence parameters were then performed on all 504,811 structures. The periodic table heatmap indicates the number of structures containing each element and is colored on a logarithmic scale.  The MatPES \rrscan dataset has similar elemental distribution (Fig.~\ref{fig:data_comp_r2scan}).
}
\end{figure}

Here, we report the launch of MatPES, an open science initiative to develop a foundational PES dataset for materials.
In addition to remedying the historical dependencies of the MPRelax set, we also improve upon the underlying DFT description of the PES. The PBE functional, predominantly used by UMLIP efforts, tends to underestimate the strength of weaker ionic and van der Waals bonds \cite{tran2016rungs}; more accurate meta-GGAs, such as the revised regularized strongly constrained and appropriately normed (\rrscan) functional \cite{furness2020r2scan}, have been developed that are better able to capture differences in local electronic bonding \cite{kingsbury2022eform} and describe intermediate van der Waals bonding without an explicit dispersion correction \cite{kothakonda2023eform}.
The initial MatPES dataset (version 2025.1) comprises accurate energies, forces, and stresses from well-converged single-point PBE and \rrscan calculations of 504,811 equilibrium and non-equilibrium structures, generated using the workflow depicted in Fig.~\ref{fig:workflow}. We demonstrate that MatPES-trained UMLIPs significantly outperform MPRelax- and OMat24-trained UMLIPs on a broad range of equilibrium, near-equilibrium and dynamic properties. This dataset is publicly available on a dedicated web site (\url{http://matpes.ai}), as well as through the Materials Project MPContribs platform \cite{huck2016mpcontribs}, and the pre-trained UMLIPs are released in the Materials Graph Library (MatGL)\cite{matgl}.

\section{Results}

\subsection{Dataset composition}

The MatPES v2025.1 dataset comprises a total of 434,712 PBE and 387,897 \rrscan calculations, with a comprehensive and relatively well-balanced coverage of all elements of the periodic table (Fig.~\ref{fig:workflow} and Fig.~\ref{fig:data_comp_r2scan}a). With the exception of inert noble gases, unstable radioactive elements, and the rare earths, each element has at least 7,000 structures. The extremely large number of structures with oxygen reflects the myriad technologically relevant oxides in the Materials Project database. There are two crucial differentiators in how the MatPES dataset was constructed that yield significant advances over previously reported datasets.

First, the structures in the MatPES dataset were sampled from an extremely large configuration space of 281 million structures and 16 billion atomic environments from 300K MD simulations of unit cells \textit{and} supercells with a pre-trained Materials 3-body Graph Network (M3GNet) UMLIP (see Methods). We found the use of supercells to be of critical importance, as they cover a wider range of atomic environments than unit cells in MD simulations (Fig.~\ref{fig:feature_space_supercell_unitcell}). Prior datasets are derived almost entirely from small unit cells due to the use of expensive DFT methods in configuration space generation.

Second, we developed an enhanced 2-stage version of the DImensionality-Reduced Encoded Clusters with sTratified (2DIRECT) sampling \cite{qiRobustTrainingMachine2024} approach to ensure data-efficient coverage of this configuration space. Briefly, each structure was encoded using a pre-trained M3GNet formation energy model\cite{chen2022m3gnet}. The intermediate output of the readout layer and the updated node features after the first graph convolution were then extracted as the structural and atomic/local environment features, respectively. Then a two-step principal component analysis (PCA) and clustering were carried out in the structural feature followed by the atomic feature space. This 2DIRECT sampling approach ensures that the MatPES dataset covers the entire space of structural and atomic environments in a data-efficient manner. The result is that the MatPES dataset is only a fraction of the size and yet samples a much larger space of structures compared to the MPtrj dataset (Fig.~\ref{fig:PCA_MatPES_vs_MPtrj}). 

\begin{figure}[htp]
    \centering
      \begin{subfigure}{0.7\columnwidth}
         \centering
         \includegraphics[width=\linewidth]{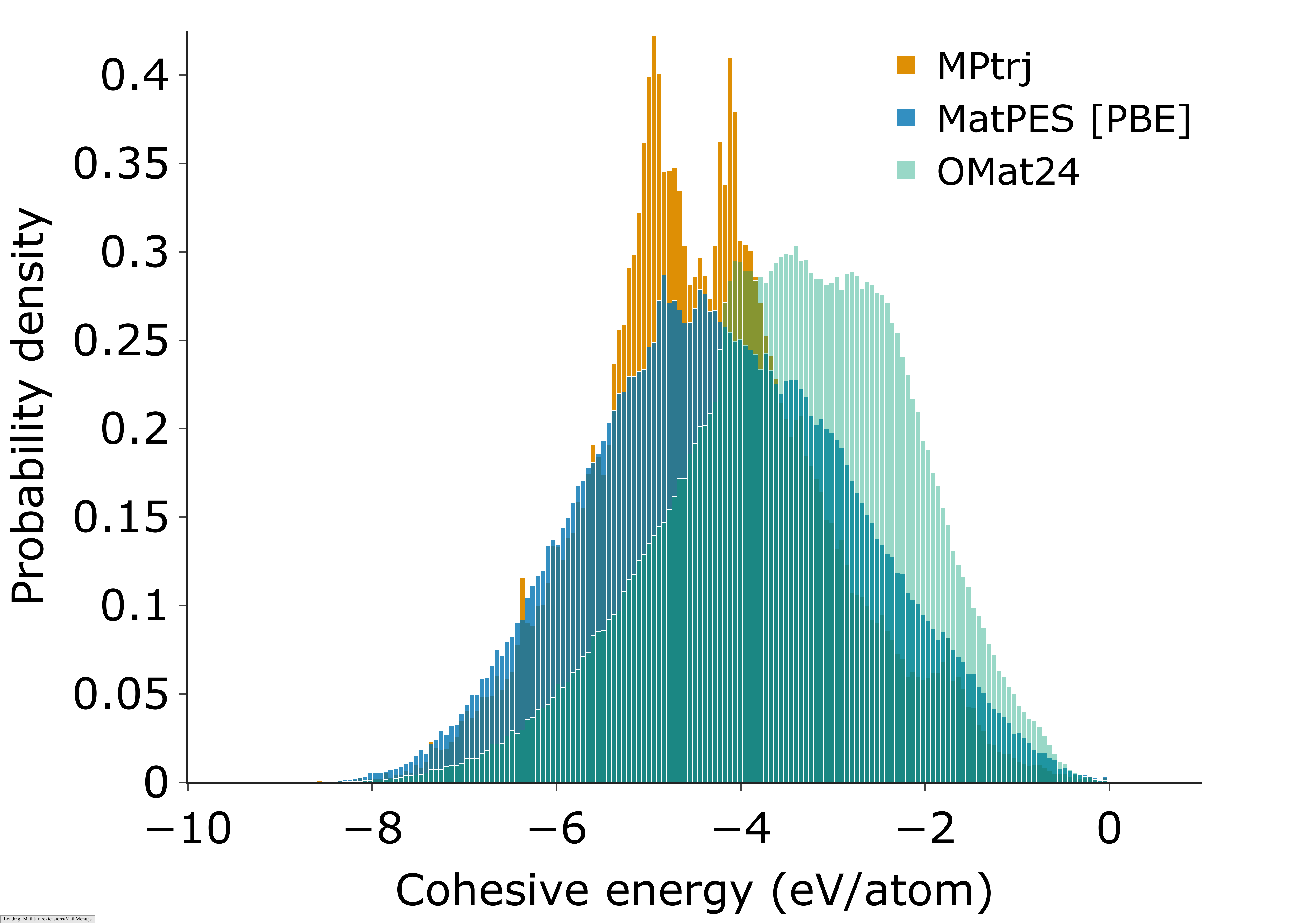}
         \caption{}
     \end{subfigure}
     \hfill
      \begin{subfigure}{0.7\columnwidth}
         \centering
         \includegraphics[width=\linewidth]{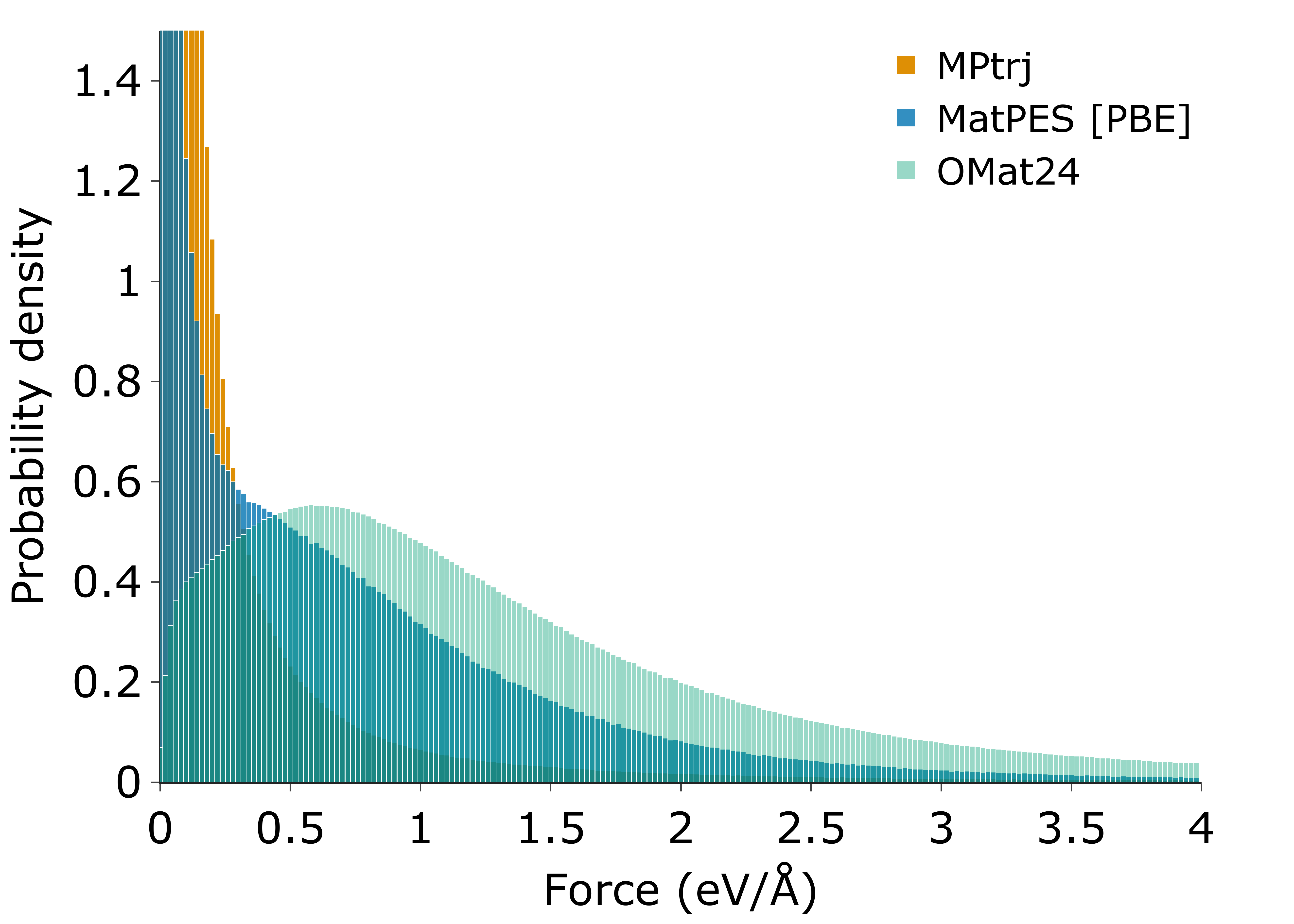}
         \caption{}
     \end{subfigure}
    \caption{
        \textbf{Coverage of the MatPES PBE dataset.} Distribution of PBE \textbf{a,} cohesive energies (\ecoh) and \textbf{b,} interatomic force magnitudes (\absforce) in the MatPES (blue), MPtrj (orange) \cite{deng2023chgnet}, and OMat24 (yellow) \cite{barrosoluque2024omat} datasets.
        The composition of the datasets are as follows: MatPES PBE: 434,712 structures (326,635 MD snapshots, 108,077 MP equilibrium structures); MPtrj: 1,580,361 structures from MP relaxations; OMat24: 1,077,382 structures. The MPtrj and OMat24 datasets contain a mixture of PBE and PBE$+U$ data, whereas MatPES PBE contains only PBE data.
    }
    \label{fig:data_comp}
\end{figure}

Compared to both MPtrj and OMat24, the MatPES PBE dataset has a more Gaussian-like distribution of cohesive energies per atom \ecoh (Fig. \ref{fig:data_comp}b) and a log-normal-like distribution of interatomic force magnitudes \absforce (Fig. \ref{fig:data_comp}c). Here, we have chosen \ecoh (Eq.~\ref{eq:cohesive_energy}) as a more appropriate measure of the overall quality of the dataset than the formation energy that is often used in the literature. \ecoh measures the stability of a solid relative to its atomic constituents and should always have a negative value except in cases of poor energy convergence or structures with unphysical bond configurations (e.g., excessively short bond distances). By including structures from 300K MD simulations, the MatPES PBE dataset samples a much wider range of \ecoh and \absforce than the MPtrj dataset. The \absforce distribution of the MPtrj dataset is especially narrow, reflecting its lack of coverage of local environments farther away from equilibrium. The OMat24 dataset has a much greater fraction of structures with higher \ecoh and \absforce, as it was constructed from hypothetical structures in the Alexandria PBE database\cite{schmidt2023alex}. Furthermore, OMat24 under-samples near-equilibrium local environments by sampling structures only from \textit{ab initio} MD (AIMD). Overall, the MatPES dataset achieves a better balance of on- and off-equilibrium structures and local environments. The MatPES \rrscan dataset has similar \ecoh and \absforce distributions as the MatPES PBE dataset (Fig. \ref{fig:data_comp_r2scan}). 

\subsection{PES benchmarks}

UMLIPs were trained using MatPES PBE and \rrscan datasets using three graph-based architectures: M3GNet,\cite{chen2022m3gnet} Crystal Hamiltonian Graph Network (CHGNet)\cite{deng2023chgnet}, and TensorNet\cite{simeon2023tensornet} (see Methods). In addition, we have trained M3GNet and CHGNet UMLIPs using the MPF and MPtrj datasets, respectively, and TensorNet UMLIPs using the MPF and OMat24 datasets to ensure a consistent basis for comparison. These architectures were selected to evaluate the performance of MatPES on both symmetry-invariant (M3GNet and CHGNet) and equivariant (TensorNet) models. All three architectures are implemented in the common Materials Graph Library (MatGL)\cite{matgl} to ensure consistency in parameter optimization. Although the authors are aware of other architectures in the literature, a comprehensive evaluation of different architectures is beyond the scope of this work. Furthermore, subsequent results will show that the differences in performance between different architectures are relatively small compared to those between different datasets. 

\begin{table}[htp]
    \centering
    \begin{tabular}{lcccc}
    UMLIP     & Energy  & Force & Stress & Magmom  \\
     \hline
     \textbf{MatPES PBE}\\
     M3GNet    & 40/45/45 & 155/177/181 & 0.734/0.898/0.888 & N/A  \\
     CHGNet    & 27/32/31  & 81/124/136 & 0.375/0.617/0.642 & 0.066/0.067/0.066\\
     TensorNet & 33/36/36 & 121/138/148 & 0.602/0.695/0.700 & N/A  \\
    \textbf{MatPES \rrscan}\\
     M3GNet    &  38/45/44 & 172/208/210 & 0.774/0.982/0.970 & N/A \\
     CHGNet    & 26/27/30  & 86/150/156 & 0.359/0.705/0.735 & 0.067/0.066/0.072 \\
     TensorNet & 32/34/34  & 139/163/163 & 0.653/0.754/0.754 & N/A \\
     \textbf{MPF}\\
     M3GNet    & 20/23/334 & 63/72/297 & 0.259/0.399/2.026 & N/A  \\
     TensorNet & 29/29/316 & 78/83/289 & 0.361/0.471/1.984 & N/A  \\
     \textbf{MPtrj}\\
     CHGNet    & 26/30/698  & 49/70/265 & 0.173/0.297/1.872 & 0.036/0.037/0.038\\
     \textbf{OMat24}\\
     TensorNet & 23/26/202 & 111/116/186 & 0.565/0.584/1.151 & N/A
    \end{tabular}
    \caption{\textbf{Mean absolute errors (MAEs) in PES quantities for trained UMLIPs.} The energies, force, stress, and magnetic moment (magmom) MAEs are reported in units of meV atom$^{-1}$, meV $\si{\angstrom}^{-1}$, GPa, and $\mu_{\mathrm{B}}$, respectively. The magmom is only used in the training of the CHGNet UMLIPs. The numbers are reported in the order of training/validation/test MAEs. The training, validation, and test sets are randomly selected from the complete dataset in proportions of 90\%:5\%:5\%, respectively. }
    \label{tab:UMLIPs_r2SCAN_MAEs}
\end{table}

The training and validation MAEs in PES quantities (energies, forces, stresses) of the MatPES UMLIPs are slightly larger than those of the MPRelax or OMat24 UMLIPs of the same architecture (Table \ref{tab:UMLIPs_r2SCAN_MAEs}). This can be attributed to the greater proportion of structures with larger forces and stresses in the MatPES dataset (Fig.~\ref{fig:data_comp}). However, the MatPES PBE UMLIPs significantly outperform the MPRelax and OMat24 trained UMLIPs in terms of the MAEs on the test set, which comprises 21,737 structures (5\%) randomly sampled from the entire MatPES dataset. The test MAEs of the MatPES PBE UMLIPs are close to the training and validation MAEs, indicating little or no overfitting. The test MAEs in energies of MPRelax and OMat24 UMLIPs are $> 4-10$ times higher than those of MatPES UMLIPs. It is not surprising that the MPRelax UMLIPs exhibit significantly higher energy errors due to the noise in the training data, as well as insufficient coverage of local environments far from equilibrium. Additionally, the exceptionally high energy error in the MPtrj CHGNet model stems from the inclusion of both PBE and PBE+$U$ calculations in its training data. The larger energy errors on the TensorNet OMat24 UMLIP is likely due to the lack of coverage of near-equilibrium local environments in OMat24. Furthermore, the TensorNet MatPES PBE UMLIP also exhibits generally uniform MAEs across all elements, while that trained on the TensorNet OMat24 UMLIP exhibits much higher errors on the rare earths and oxygen (Fig.~\ref{fig:mae-el-heatmap}).

\subsection{Property benchmarks}

To evaluate UMLIPs with different architectures and/or training data, we developed a set of equilibrium (relaxed structure similarity, formation energy), near-equilibrium (bulk and shear moduli, constant-volume heat capacity, force softening) and molecular dynamics (stability, ionic conductivity, efficiency) property benchmarks, collectively referred to as MatCalc\cite{matcalc}-Bench. To ensure unbiased evaluation of UMLIPs trained with different datasets, the benchmark test data were curated from independent sources, including the Materials Project\cite{jain2013mp,dejong2013elastic}, Alexandria,\cite{loewphonon2024} WBM\cite{wangpredicting2021}, Graph Networks for Materials Science (GNoME)\cite{merchant2023gnome}, WBM high energy states\cite{deng2024soft}, Materials Virtual Lab databases, summarized in Tab.~\ref{tab:benchmarking_tasks}.

\subsubsection{Equilibrium benchmarks}

\begin{figure}[htp]

    \centering
      \begin{subfigure}{0.45\columnwidth}
         \centering
         \includegraphics[width=\linewidth]{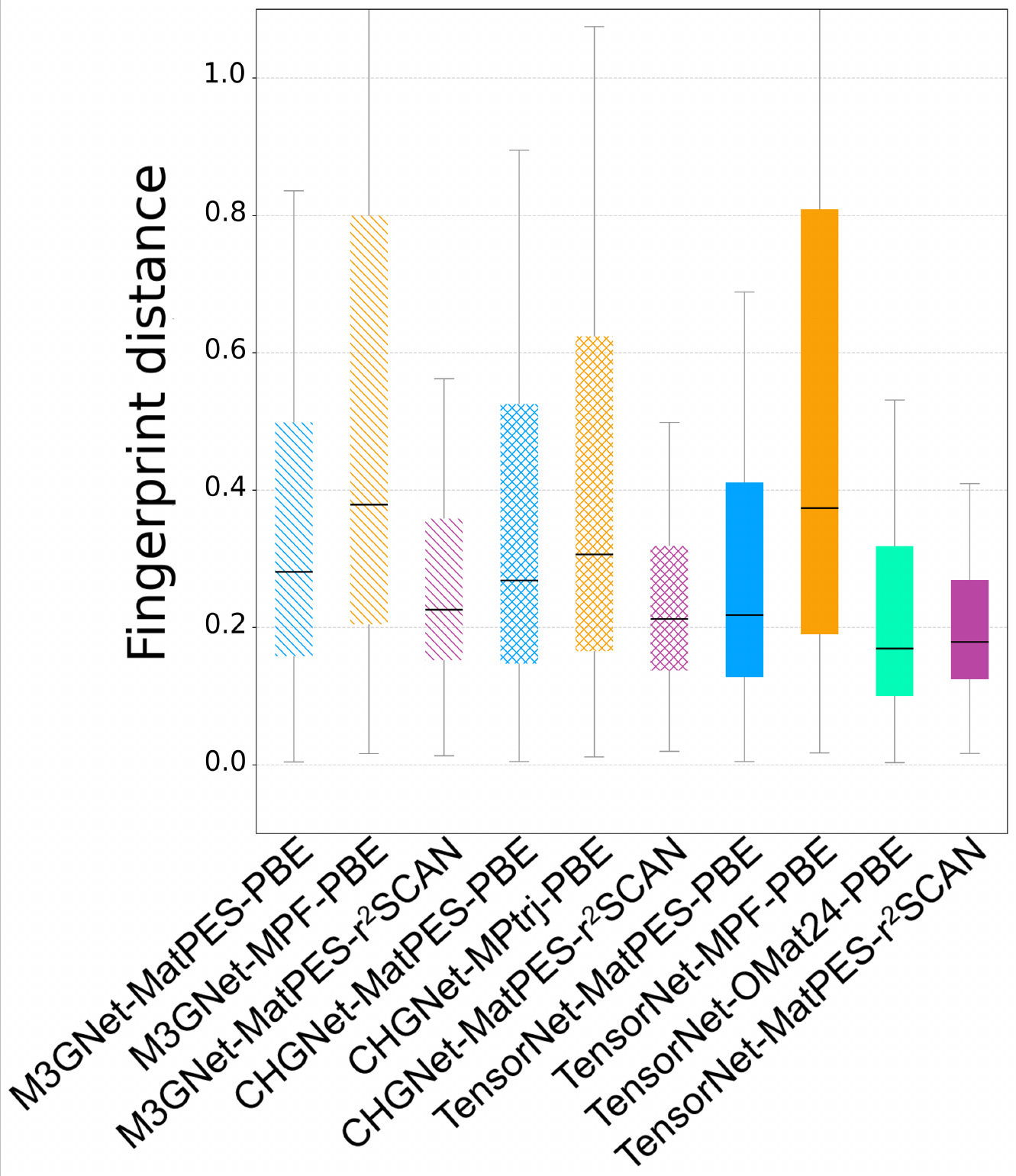}
         \caption{}
     \end{subfigure}
     \hfill
      \begin{subfigure}{0.45\columnwidth}
         \centering
         \includegraphics[width=\linewidth]{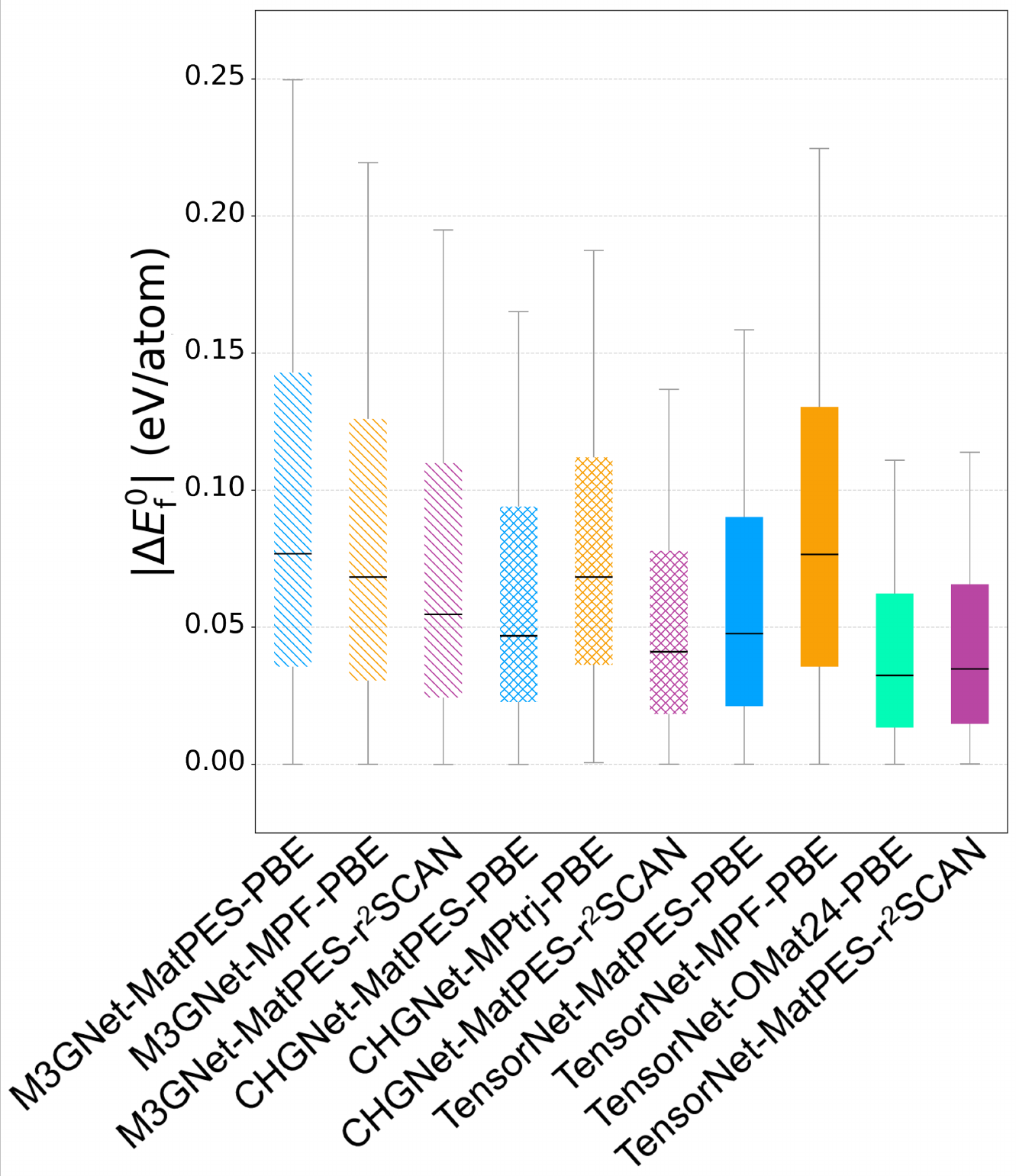}
         \caption{}
     \end{subfigure}
\caption{
    \label{fig:equilibrium}\textbf{Evaluation of UMLIPs on equilibrium properties.} Distribution of the \textbf{a,} structural similarity fingerprint distance and \textbf{b,} formation energy per atom error between UMLIP and DFT-relaxed structures with the  PBE and \rrscan functionals. A random direction perturbation was applied to all sites of 1,000 out-of-domain PBE-relaxed and \rrscan-relaxed structures randomly sampled from the WBM\cite{wangpredicting2021} and GNoME\cite{merchant2023gnome} databases, respectively, prior to geometry optimization using UMLIPs. CrystalNN\cite{zimmerman2020crystalnn} was used to compute the fingerprint distance (see Methods).
}
\end{figure}

As shown in Fig.~\ref{fig:equilibrium}, MatPES UMLIPs generally outperform MPRelax UMLIPs of the same architecture for equilibrium properties such as structural relaxations, but predict formation energies with comparable or slightly lower accuracy. Structures relaxed using MatPES PBE UMLIPs tend to have a lower mean ``fingerprint'' distance (defined in ``Methods'') from DFT-PBE relaxed structures and with a lower variance than those relaxed using MPRelax UMLIPs. The MAE in formation energy per atom for the MatPES-trained M3GNet UMLIP is slightly higher than that of the MPF-trained M3GNet UMLIP. The MatPES CHGNet UMLIP, however, performs better than the MPtrj CHGNet UMLIP. We believe this is because the CHGNet model used here has a much higher model complexity (2,700,000 parameters) than the M3GNet model (664,000 parameters) and thus is able to better learn the diverse PES landscape of the dataset. The equivariant TensorNet models are able to achieve slightly lower fingerprint distance and formation energy errors than the invariant M3GNet and CHGNet models, despite having a relatively small number of parameters (838,000). The performance of MatPES \rrscan UMLIPs is also generally excellent and similar to MatPES PBE UMLIPs. 

\subsubsection{Near-equilibrium benchmarks}

\begin{figure}[htp]
\centering
\includegraphics[width=1\columnwidth]{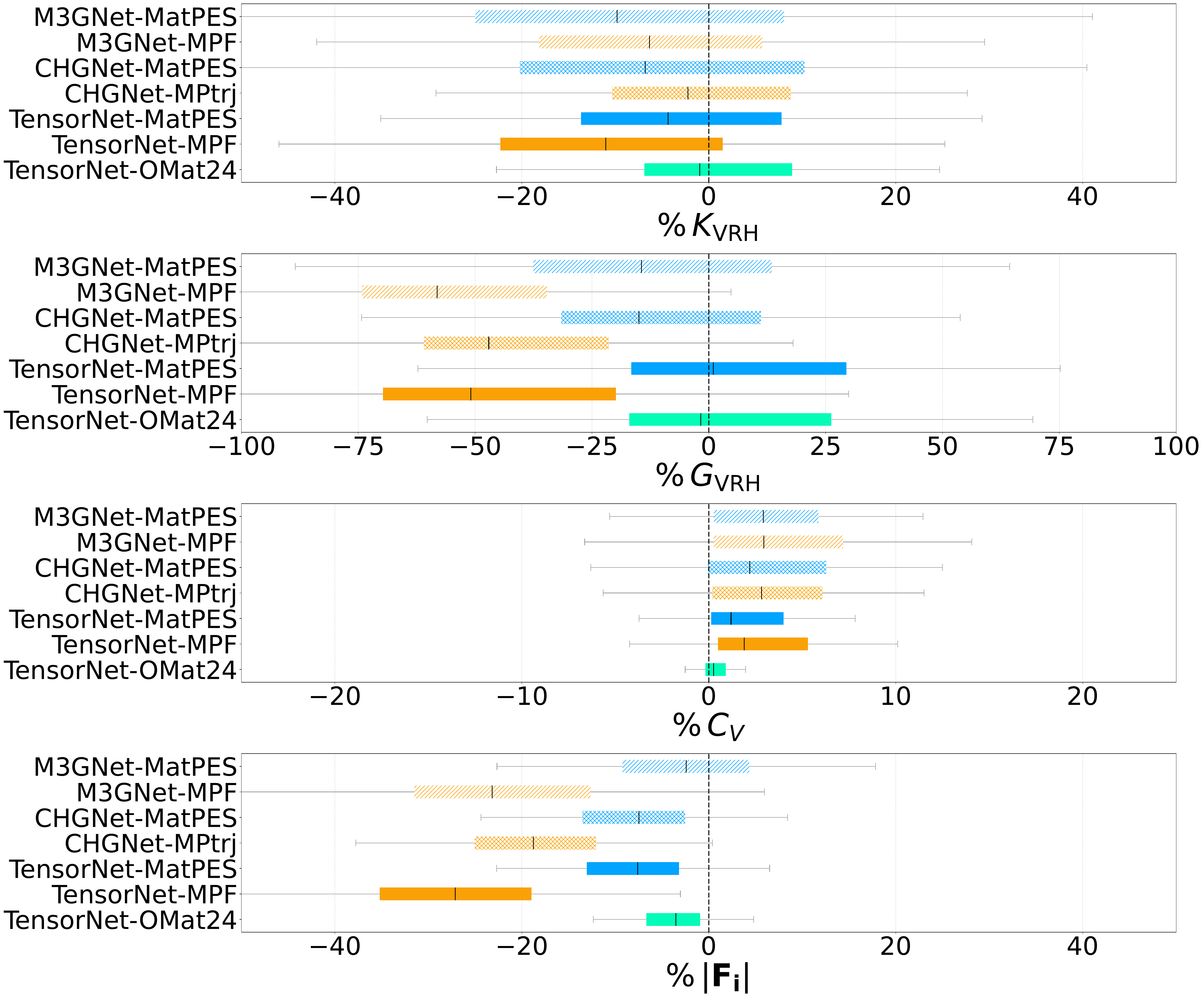}
\caption{
\label{fig:near_equilibrium}\textbf{Evaluation of UMLIPs on near-equilibrium properties.} Distribution of the percentage errors in the predicted \textbf{a,} bulk moduli ($K_{VRH}$), \textbf{b,} shear moduli ($G_{VRH}$), \textbf{c,} constant-volume heat capacities ($C_V$), and \textbf{d,} off-equilibrium forces ($|\mathbf{F_i}|$) of MatPES PBE, MPRelax and OMat24 UMLIPs compared to the DFT ground truth.
The elastic moduli benchmarks comprises 3,959 binary compounds with computed elastic moduli in the Materials Project.\cite{jain2013mp,dejong2013elastic} The $C_V$ benchmark is derived within the harmonic approximation using 1,170 structures from the Alexandria phonon database\cite{loewphonon2024}. The $|\mathbf{F_i}|$ benchmark is computed from all 979 configurations in the WBM high energy states database.\cite{wangpredicting2021}
    }
\end{figure}

The evaluation of UMLIPs on near-equilibrium properties was carried out primarily on MatPES PBE trained UMLIPs due to the lack of large \rrscan datasets in the literature. Compared to MPRelax and OMat24 UMLIPs, we find that MatPES UMLIPs generally yield significant improvements in the prediction of the shear modulus $G_\mathrm{VRH}$ and off-equilibrium forces $|\mathbf{F_i}|$, while having similar performance in the prediction of the bulk modulus $K_\mathrm{VRH}$ and constant-volume heat capacity $C_V$ (Fig. \ref{fig:near_equilibrium}). The slightly better performance of MPRelax UMLIPs on $K_\mathrm{VRH}$ is expected due to the inclusion of a greater fraction of near-equilibrium relaxation structures in the dataset. The TensorNet-OMat24 UMLIP provides the most accurate predictions of $C_V$, probably due to the inclusion of numerous rattled configurations through Boltzmann sampling. MatPES UMLIPs largely correct the systematic under-prediction of the PES curvature by MPRelax UMLIPs.\cite{deng2024soft} We believe the remanent small systematic under-estimation of PES curvatures by MatPES UMLIPs will be addressed with the addition of structures from higher-temperature MD simulations in future MatPES dataset releases.

\subsubsection{Molecular dynamics benchmarks}

A primary application of UMLIPs is in MD simulations, but most existing benchmarks often do not include an assessment of UMLIP performance on MD stability or properties due to the lack of reference ab initio data. Here, a database of AIMD simulations of 172 battery materials performed by the Materials Virtual Lab over the past decade (MVL-batt) is used to evaluate UMLIPs. 

\begin{figure}[htp]
\centering
\begin{subfigure}{\columnwidth}
    \includegraphics[width=\columnwidth]{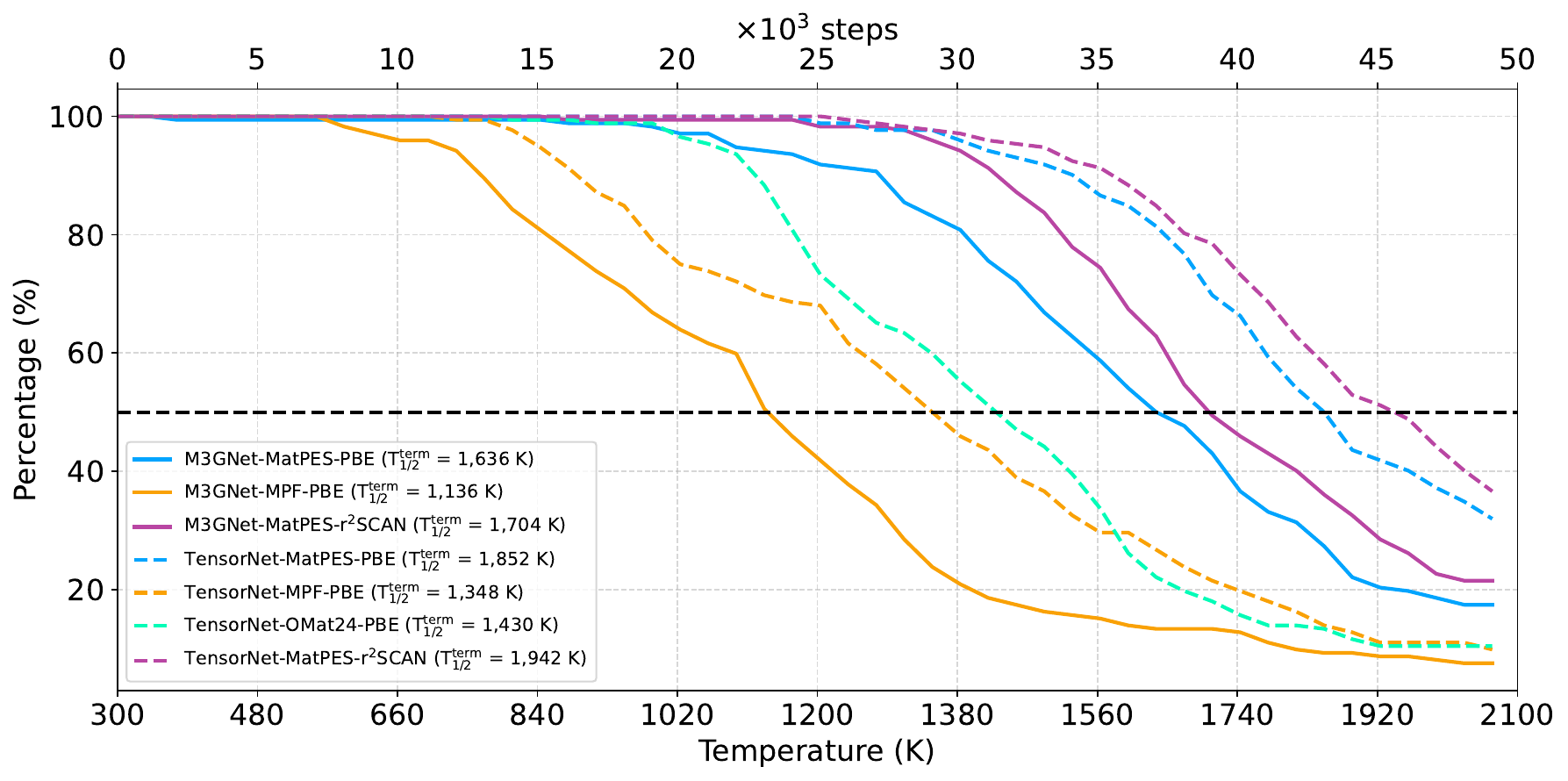}
    \caption{}
\end{subfigure}
\begin{subfigure}{\columnwidth}
    \includegraphics[width=\columnwidth]{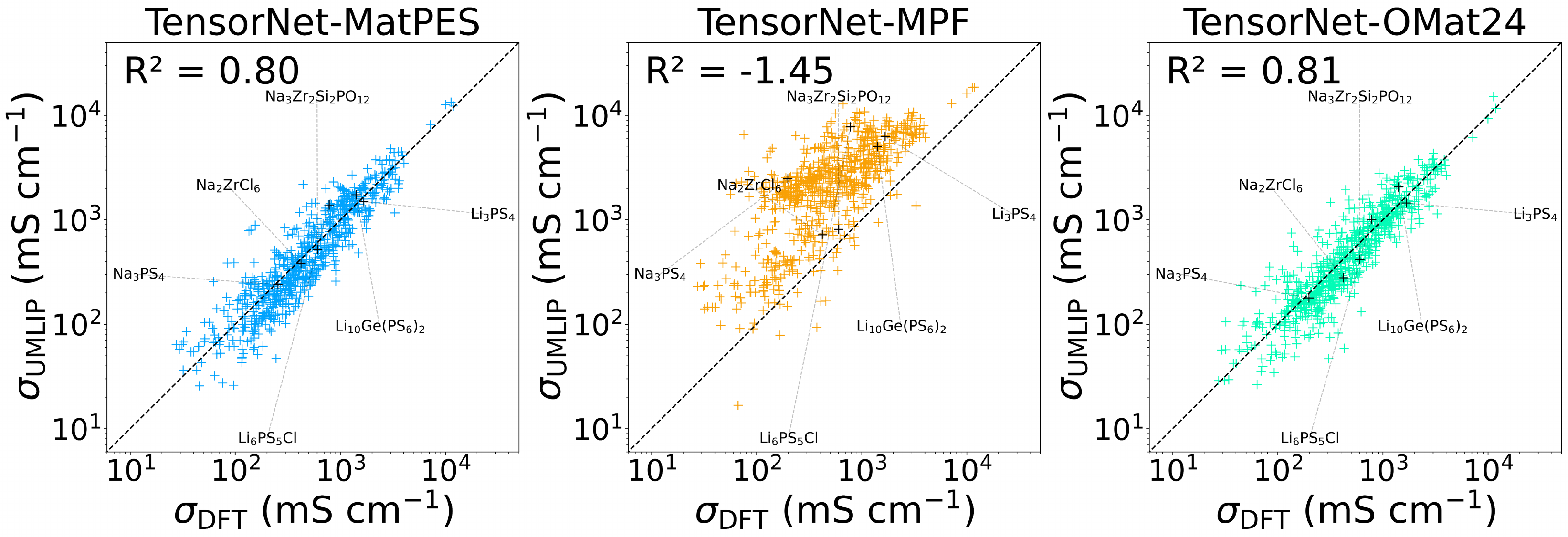}
    \caption{}
\end{subfigure}
\caption{
    \label{fig:md_properties}\textbf{Evaluation of UMLIPs on molecular dynamics (MD) properties of the MVL-Batt test set of 172 Li and Na-containing battery materials.} \textbf{a,} Distributions of the MD termination steps of UMLIPs based a controlled heating protocol from 300 K to 2,100 K at 1 bar over 50 ps with a 1 fs time step for the MVL-Batt test set. Simulations terminate due to volume explosion ($V_t \geq 1.5V_0$) or atom loss. Three runs were performed per model for statistical reliability. Only the M3GNet and TensorNet architectures were used for these simulations. The metric to assess MD stability is the median termination temperature $T_{1/2}^{term}$, indicated for each of the UMLIPs in the legend. \textbf{b,} Parity plots of the UMLIP-predicted ($\sigma_{\mathrm{MLIP}}$) against the AIMD ($\sigma_{\mathrm{DFT}}$) Li/Na ionic conductivities of the MVL-Batt test set. A total of 698 \textit{NVT} MD simulations at multiple temperatures (300-2,100 K) were performed. The data points for six well-known Li and Na solid electrolyte materials at 1,000 K are labeled for reference. The $R^2$ score is calculated from the mean squared error in $\mathrm{log}(\sigma)$ to ensure a robust evaluation across multiple orders of magnitude. 
}
\end{figure}

A basic requirement for MD simulations is stability, which we assessed using the median termination temperature $T_{1/2}^{\mathrm{term}}$ of 300 K-2,100 K linear heating MD simulations of the MVL-batt test structures   (Fig.~\ref{fig:md_properties}a). Common causes of terminations in MD simulations are volume explosion and atom loss. Given the same architecture, MatPES PBE UMLIPs exhibit significantly better MD stability, i.e., much larger $T_{1/2}^{\mathrm{term}}$, compared to MPRelax and OMat24 UMLIPs. By 1500 K, less than 10\% of the TensorNet-MatPES-PBE and TensorNet-MatPES-\rrscan simulations have terminated, while about 55\% and 65\% of TensorNet-OMat24-PBE and TensorNet-MPF-PBE simulations, respectively, have terminated. Equivariant TensorNet UMLIPs generally exhibit better stability than invariant M3GNet UMLIPs. Also, MatPES \rrscan UMLIPs exhibit better stability than MatPES PBE UMLIPs, which is likely due to the improved description of interatomic bonding by the \rrscan functional.

The TensorNet-MatPES UMLIP significantly outperforms the TensorNet-MPF UMLIP (Fig. \ref{fig:md_properties}b) in terms of the predicted ionic conductivities ($\sigma_{\mathrm{MLIP}}$).  TensorNet-MPF UMLIP significantly overestimates $\sigma_{\mathrm{MLIP}}$ and has large errors spanning orders of magnitude (negative $R^2$). Again, this is likely due to the lack of off-equilibrium structures in the MPF dataset. The performance of the TensorNet-OMat24 UMLIP is comparable to that of the TensorNet-MatPES UMLIP, but with a training data size that is 250 times larger.

\section{Discussion}

In recent years, advances in UMLIPs have been driven by ever larger models with increasing numbers of parameters, with the most performant models trained on closed source industry datasets, and evaluated on a narrow set of properties, mainly formation energies and stability classification.\cite{riebesell2024mbd} We believe that this trend presents significant risks. Reliance on large, proprietary datasets exacerbates reproducibility challenges and creates barriers that limit wider scientific participation. For instance, the training time for TensorNet on the MatPES-PBE dataset ($\sim 400,000$ structures) is about 15 minutes per epoch on a single Nvidia RTX A6000 GPU, while that for the same model on the OMat24 dataset ($\sim 100$ million structures) is around 20 hours per epoch with sixteen Nvidia A100 GPUs. Furthermore, larger models require greater computational resources to run, restricting their feasibility for large-scale simulations (Tab.~\ref{tab:speed_differences}). The overemphasis on formation energies and stability classification overlooks other critical material properties essential for real-world applications.

Our findings challenge the notion that larger is always better for PES datasets. UMLIPs trained on the well-sampled MatPES dataset ($\sim$ 400,000 structures) perform as well as or better than those trained on much larger datasets, such as MPtrj ($\sim$1 million structures) and OMat24 ($\sim$100 million structures). By introducing the first \rrscan dataset that spans the periodic table, we also address a critical gap in PES descriptions from higher-order DFT methods. Finally, the MatCalc benchmark\cite{matcalc} provides a comprehensive evaluation of UMLIPs across a wide range of PES-derived properties.

This dataset release marks the beginning of the MatPES initiative. There is undoubtedly room to further expand the MatPES dataset beyond 300 K MD-sampled crystals, for example, by incorporating higher-temperature/pressure MD snapshots, defect structures, hypothetical materials, surfaces and interfaces, transition states, etc. The 2DIRECT workflow developed in this work provides a robust approach to these potential augmentation efforts in a data-efficient manner. We anticipate that these efforts will further enhance the reliability and accuracy of UMLIPs across diverse applications, solidifying their role as a cornerstone for materials discovery and design.

\section{Methods}

\subsection*{Configuration space generation and sampling}

Of the 154,718 ground-state structures in Materials Project (v2022.10.28), a total of 151,768 structures with $< 250$ atoms in the unit cell were selected. For each structure, supercells were constructed so that the minimum distance between periodic neighbors is more than 7.5 \AA{}. Excluding supercells with more than 250 atoms and duplicates, the final set of initial structures totals 281,572. \textit{NpT} MD simulations were then performed on these initial structures for 50,000 time steps at 300 K and 1 atm with the pre-trained M3GNet-MP-2021.2.8-DIRECT UMLIP\cite{qiRobustTrainingMachine2024} implemented in the Materials Graph Library (\texttt{MatGL})\cite{matgl} and Large-scale Atomic/Molecular Massively Parallel Simulator (LAMMPS) \cite{thompson2022lammps}.
The MD time interval was set to 2 fs and 0.5 fs for structures without and with hydrogen, respectively. In each MD run, 1,001 structures were dumped, resulting in 281,853,572 MD snapshot structures. 
 
The M3GNet formation energy model\cite{chen2022m3gnet} was used to encode the 281 million MD structures and their $>$16 billion atomic environments. The structural and atomic features had dimensionalities of 128 and 64, respectively. 
First, the 281 million structures were clustered into 15,192 clusters according to their locations in the structural feature space. Here, the first 16 PCs were used as structural features, as determined by Kaiser's rule.\cite{qiRobustTrainingMachine2024} The threshold of BIRCH clustering was set to 0.5, as determined by memory limitations. Second, in each cluster, BIRCH clustering was performed in the atomic feature space with a threshold of 0.1. Normalization and dimensionality reduction were carried out to transform the 64-D M3GNet atomic features to 8-D vectors. The scaler and PCA were fitted to the atomic features of the 154,718 ground-state structures only, as fitting them directly with the 16 billion atomic features is untractable. Finally, the structure with the smallest number of atoms in each cluster was selected to minimize DFT computational costs. 

\subsection*{Density functional theory calculations}

The MatPES training data were obtained from single-point (static) calculations using the Vienna \textit{ab initio} Simulation Package (VASP) \cite{kresse1993vasp1,kresse1994vasp2,kresse1996vasp3,kresse1996vasp4} version 6.4.x. The VASP input parameters were carefully benchmarked for energy and force convergence and implemented as a ``MatPESStaticSet'' class in the open-source Python Materials Genomics (\texttt{pymatgen}) library\cite{ong2013pymagtgen}, versions \texttt{2025.1.9} and newer. The public availability of ``MatPESStaticSet'' allows users to generate additional training data to fine-tune MLIPs in a consistent manner. Tables \ref{tab:matpes_incar} and \ref{tab:matpes_potcar} provide MatPES-compatible \texttt{INCAR} (including $k$-point density) and \texttt{POTCAR} settings, respectively. For large-scale data generation, the \texttt{MatPESStaticFlowMaker} function in the \texttt{atomate2} workflow orchestration package\cite{ganose2025atomate2} can be used.

Both the Perdew-Burke-Ernzerhof (PBE) \cite{perdew1996pbe} generalized gradient approximation (GGA) and \rrscan meta-GGA \cite{furness2020r2scan} were employed to approximate the exchange-correlation energy.
The self-consistent PBE orbitals were used as initial orbitals to accelerate \rrscan calculations, useful for structures, which are far from equilibrium or with challenging bond arrangements.
The most recent ``PBE 64'' pseudopotential library from VASP based on PBE all-electron calculations was used, and Gaussian Fermi surface broadening was used to ensure that interatomic forces in metals do not suffer from known errors in the tetrahedron method \cite{dossantos2023smear}.
Because Gaussian smearing contributes a small error to the total energy and forces via the electronic pseudo-entropy \cite{dossantos2023smear}, we have ensured that the pseudo-entropy term contributes less than 1 meV atom$^{-1}$ to the total free energy by improving the ``LargeSigmaHandler'' in the \texttt{custodian} python package.
This handler dynamically checks the pseudo-entropy term during a VASP calculation, and decreases the width of Fermi surface broadening if the pseudo-entropy exceeds 1 meV/atom.
The total DFT energy extrapolated to zero electronic smearing, the interatomic (Hellmann-Feynman) forces, and symmetric stress tensor were then used to train MLIPs.
The lower success rate of the \rrscan (77\%) compared to the PBE (86\%) calculations is consistent with the lower stability and higher cost of the \rrscan functional \cite{mejiarodriguez2020r2scanl,kingsbury2022eform}. A fixed amount of computational resources was allocated to this project. Thus, the definition of ``successful'' calculations are those that ran within the budgeted computational resources \emph{and} that converged, while  ``failed'' calculations either did not complete within budgeted resources \emph{or} did not converge.

\subsection*{UMLIP training}

All ULIPs were trained using the \texttt{MatGL} package \cite{matgl}, version 1.13. The key training hyperparameters are summarized in Table
\ref{tab:hyperparameters}. All other hyperparameters were set to their default values. The total learnable parameters of M3GNet\cite{chen2022m3gnet}, TensorNet\cite{simeon2023tensornet}, and CHGNet\cite{deng2023chgnet} are 664,000, 838,000, and 2,700,000, respectively. 

\begin{table}[htp]
    \centering
    \begin{tabular}{p{4cm}|ccc}
        Parameter & M3GNet & TensorNet & CHGNet \\
        \hline
        Loss function & $L_1$ & $L_1$ & Huber with $\delta=0.1$\\
        Energy weight & 1.0 & 1.0 & 1.0\\
        Force weight & 1.0 & 1.0 & 1.0 \\
        Stress weight & 0.1 & 0.1 & 0.05\\
        Magmom weight & N.A. & N.A. & 1.0\\
        Optimizer & AMSGrad AdamW \cite{loshchilov2019adamw} & AMSGrad AdamW \cite{loshchilov2019adamw} & Adam \cite{Kingma_Ba_2014}\\
        Initial learning rate &  $10^{-3}$ & $10^{-3}$ & $10^{-3}$\\
        Min. learning rate & $10^{-5}$ & $10^{-5}$ & $10^{-5}$\\
        Bond distance cutoff & 5 \AA{} & 5 \AA{} & 6 \AA{}\\
        Three-body cutoff & 4 \AA{} & N.A & 3 \AA{} \\ 
        Number of hidden neurons & 128 & 128 & 128\\
        Number of graph convolution layers & 3 & 2 & 5\\
        Basis expansion order & $\texttt{max\_n}=3, \texttt{max\_{l}}=3$ & $\texttt{max\_n}=3$ & $\texttt{max\_n}=63, \texttt{max\_f}=32$
    \end{tabular}
    \caption{Training hyperparameters for UMLIPs}
    \label{tab:hyperparameters}
\end{table}

\subsection*{Benchmarking metrics}

\subsubsection{Structure optimization}

We randomly selected 1,000 structures relaxed with PBE from the WBM database \cite{wangpredicting2021} and 1,000 relaxed with \rrscan{} from the MP recompute of the GNoME materials \cite{merchant2023gnome}.
Random atomic displacements of 0.1 \AA{} were applied to these structures and then UMLIPs were used to relax the perturbed structures with the FIRE optimizer \cite{bitzek2006fire} and a 0.05 eV/\AA{} force convergence criterion.
The CrystalNN method\cite{zimmerman2020crystalnn} was used to compute the ``fingerprint'' vector of a structure based on its local environments. The similarity of a UMLIP-relaxed structure to its reference DFT-relaxed structure is then the Euclidean distance between their corresponding fingerprint vectors, with a lower distance indicating greater similarity between structures.

\subsubsection{Cohesive and formation energy}

The cohesive energy $ E_\mathrm{coh}$ of a solid is defined as its total energy $E_\mathrm{solid}$ relative to its atomic (gas phase), $E^\mathrm{atom}_i$, constituents with stoichiometric coefficients $N_i$,
\begin{equation}
    \ecoh = \frac{1}{\sum_i N_i } \left[ E_\mathrm{solid} - \sum_i N_i E^\mathrm{atom}_i \right].
    \label{eq:cohesive_energy}
\end{equation}
The formation energy of a solid at 0 K, \eform, is defined as its total energy, $E_\mathrm{solid}$, relative to the stoichiometry-weighted energies of the 0K ground state (solid phase) of its elemental constituents, $E_i$:
\begin{equation}
     \eform = \frac{1}{\sum_i N_i } \left[ E_\mathrm{solid} - \sum_i N_i E_i \right].
    \label{eq:formation_enthalpy}
\end{equation}
Both the cohesive and formation energies are normalized by the number of atoms in the structure.


\subsubsection{Elastic moduli and heat capacity}

The bulk ($K_\mathrm{VRH}$) and shear ($G_\mathrm{VRH}$) moduli were extracted from the elastic tensor $\mathbf{C}$ using the Voigt-Reuss-Hill (VRH) averaging scheme\cite{hillvrh1952}.

We selected the 3,959 binary structures from the Materials Project \cite{jain2013mp,dejong2013elastic} with converged PBE elastic tensors. These structures were relaxed using UMLIPs and then subjected to a series of normal and shear deformations to determine these constants. Linear strain values of (\(-0.01, -0.005, 0.005, 0.01\)) and (\(-0.06, -0.03, 0.03, 0.06\)) were used for each normal and shear modes, respectively. Shear deformation generally induced a weaker elastic response compared to normal deformation, so larger strains were required to ensure an accurate stress-strain fit.

The constant-volume heat capacity $C_V$ is obtained from the partial derivative of the vibrational internal energy $U_\mathrm{vib}$ with respect to temperature under the harmonic phonon approximation:
\begin{equation}
C_V(T) = \left( \frac{\partial U_\mathrm{vib}(T)}{\partial T} \right)_V.
\label{eq:heat_capacity}
\end{equation}
Here, we focused on the heat capacity at room temperature ($T$ = 300 K). As references, all 1,170 binary materials with converged PBE calculations in the Alexandria phonon database \cite{loewphonon2024}, were collected and relaxed by UMLIPs. A \( 2 \times 2 \times 2 \) supercell was generated for each UMLIP-relaxed structure and an atomic displacement of 0.015 \text{\AA} was introduced into the supercell. Subsequently, the interatomic forces were calculated using UMLIPs for each supercell with displacement. The force constants were then obtained from the displacements and forces.

The dynamical matrix was constructed for each phonon wave vector $\mathbf{q}$ in the Brillouin zone using the force constants obtained previously. The matrix was then diagonalized to obtain the phonon frequencies $\omega(\mathbf{q}\nu)$ and their corresponding eigenvectors. The Brillouin zone was sampled using a $\Gamma$-centered mesh with density proportional to the reciprocal lattice vector, scaled by 100. For each $q$-point, all phonon modes were calculated and the resulting frequencies $\omega(\mathbf{q}\nu)$ were used to calculate $U_\mathrm{vib}$ at a given temperature $T$ as
\begin{equation}
U_\mathrm{vib}(T) = \sum_{\mathbf{q}\nu} \hbar \,\omega(\mathbf{q}\nu) 
\left\{ \frac{1}{2} + \left[ \exp \left(\frac{\hbar \,\omega(\mathbf{q}\nu)}{k_\mathrm{B} \,T} \right) - 1 \right]^{-1} \right\},
\label{internal_energy}
\end{equation}
where $k_\mathrm{B}$ and $\hbar$ denote the Boltzmann constant and the reduced Planck constant, respectively.

\subsubsection{MD benchmarks}

The MD stability of a UMLIP is assessed using the median time-step of failure in a heating MD simulation in LAMMPS. A database of AIMD simulations of 172 battery materials performed by the Materials Virtual Lab over the past decade (MVL-batt) was used. Each material was relaxed and then heated from 300 K to 2,100 K at 1 bar over 50,000 time steps of 1 fs (50 ps in total). The final timestep at which the simulation crashed, due to explosion of the cell ($V_t \geq 1.5V_0$) or loss of atoms, was recorded for each material. Three runs were carried out for each material under the same conditions. 

To compute the ionic conductivity,\textit{NVT} MD simulations were performed using DFT and UMLIPs. A total of 698 simulations were carried out across a temperature range from 300 K to 2,100 K for the MVL-batt materials. Each simulation was run for at least 110 ps. The first 10 ps of the simulation were for equilibration, and the remaining 100 ps was used to compute the mean square displacement (MSD) of the diffusing species. The ionic conductivity $\sigma$ is derived from the Nernst-Einstein equation:

\begin{equation}
\sigma = \frac{z^2 F^2 \rho}{R \, T} \, D,
\label{eq:ionic_conductivity}
\end{equation}

where $D$ is the diffusivity, $\rho$ is the number density of diffusing ions, $T$ is the absolute temperature, $z$ is the ionic charge of the diffusing species ($z$ = 1), $F$ is Faraday's constant, and $R$ is the universal gas constant. $D$ is obtained from a linear fit of the MSD versus time, according to the Einstein relation:

\begin{equation}
D = \lim_{\Delta t \to \infty} \frac{1}{2d \Delta t} 
\Big\langle \big| \mathbf{r}_i(t + \Delta t) - \mathbf{r}_i(t) \big|^2 \Big\rangle_{i,t},
\label{eq:diffusion_coefficient}
\end{equation}

where $\Delta t$ denotes the time interval over which the particle displacement is measured and $\mathbf{r}_i(t + \Delta t)$ represents the position vector of the $i^\mathrm{th}$ diffusing ion at time $t + \Delta t$. $d=3$ is the dimensionality of the system, and $\Big\langle \cdots \Big\rangle$ indicates an ensemble average over the diffusing ion and time.

\section*{Data Availability}

The MatPES dataset, with associated metadata and usage guide, is available via the MatPES.ai website (http://matpes.ai). The dataset is also available on the MPContribs platform \cite{huck2016mpcontribs} at \url{https://materialsproject-contribs.s3.amazonaws.com/index.html#MatPES_2025_1/} as a bulk download, and via the explorer at \url{https://next-gen.materialsproject.org/contribs/projects/MatPES_2025_1}.

\section*{Code Availability}

All software used in this work are publicly available in open-source libraries. The UMLIP architectures are available in the Materials Graph Library (\texttt{MatGL}).\cite{matgl} The MatCalc-Bench is implemented in the \texttt{MatCalc} library.\cite{matcalc} The DFT computation and analysis code are available in the \texttt{pymatgen}\cite{ong2013pymagtgen}, \texttt{atomate2} \cite{ganose2025atomate2}, and \texttt{emmet-core} libraries.

\bibliography{references}

\begin{acknowledgement}

This work was intellectually led by the U.S. Department of Energy, Office of Science, Office of Basic Energy Sciences, Materials Sciences and Engineering Division under contract No. DE-AC02-05-CH11231 (Materials Project program KC23MP).
This research used resources of the National Energy Research Scientific Computing Center (NERSC), a Department of Energy Office of Science User Facility using NERSC award DOE-ERCAP0026371.
A portion of the research was performed using computational resources sponsored by the Department of Energy's Office of Energy Efficiency and Renewable Energy and located at the National Renewable Energy Laboratory.
A portion of the research used the Lawrencium computational cluster resource provided by the IT Division at the Lawrence Berkeley National Laboratory (Supported by the Director, Office of Science, Office of Basic Energy Sciences, of the U.S. Department of Energy under Contract No. DE-AC02-05CH11231)
T. W. Ko also acknowledges the support of the Eric and Wendy Schmidt AI in Science Postdoctoral Fellowship, a Schmidt Futures program.
We acknowledge Xiaoxu Ruan for initial discussions regarding the \texttt{MatPESStaticSet}.

\end{acknowledgement}

\section{Author Contributions}

A.D.K.: benchmarking and performing DFT calculations; revision of DFT workflows; selecting equilibrium structures to augment MD set from the Materials Project and to perform \rrscan benchmarks from the MP-GNoME structures; writing and editing the manuscript; design of the MPContribs dataset.
R.L.: design and implementation of the benchmarking workflows; performing the equilibrium properties, elastic moduli, constant-volume heat capacity and molecular dynamics properties benchmarks; collating the AIMD results; writing and editing the manuscript.
J.Q.: project initiation and conception; design and execution of the configuration space expansion and 2DIRECT sampling; development of batched dataset loading in MatGL and training of TensorNet-OMat24; development of \texttt{MatPESStaticSet}; manuscript writing.
T.W.K.: Training M3GNet and TensorNet models for both MPF and MatPES datasets; manuscript editing.
B.D.: Training CHGNet; benchmarking force softening; manuscript editing.
J.R.: conceptualization of sampling method; benchmarking of DFT parameters; implementation of \texttt{MatPESStaticSet}; writing workflows; manuscript editing.
G.C.: manuscript editing; provision of computational resources.
K.A.P.: project design; manuscript editing; provision of computational and data dissemination resources.
S.P.O.: project design and conception; manuscript writing; provision of computational and data dissemination resources; \url{matpes.ai} website development.

\renewcommand{\thepage}{S\arabic{page}}
\renewcommand{\thesection}{S\arabic{section}}
\renewcommand{\theequation}{S\arabic{equation}}
\renewcommand{\thetable}{S\arabic{table}}
\renewcommand{\thefigure}{S\arabic{figure}}

\clearpage

\section*{Supplementary Information: A Foundational Potential Energy Surface Dataset for Materials}

\begin{figure}[htp]
    \centering
    \begin{subfigure}{0.9\columnwidth}
        \centering
        \includegraphics[width=\columnwidth]{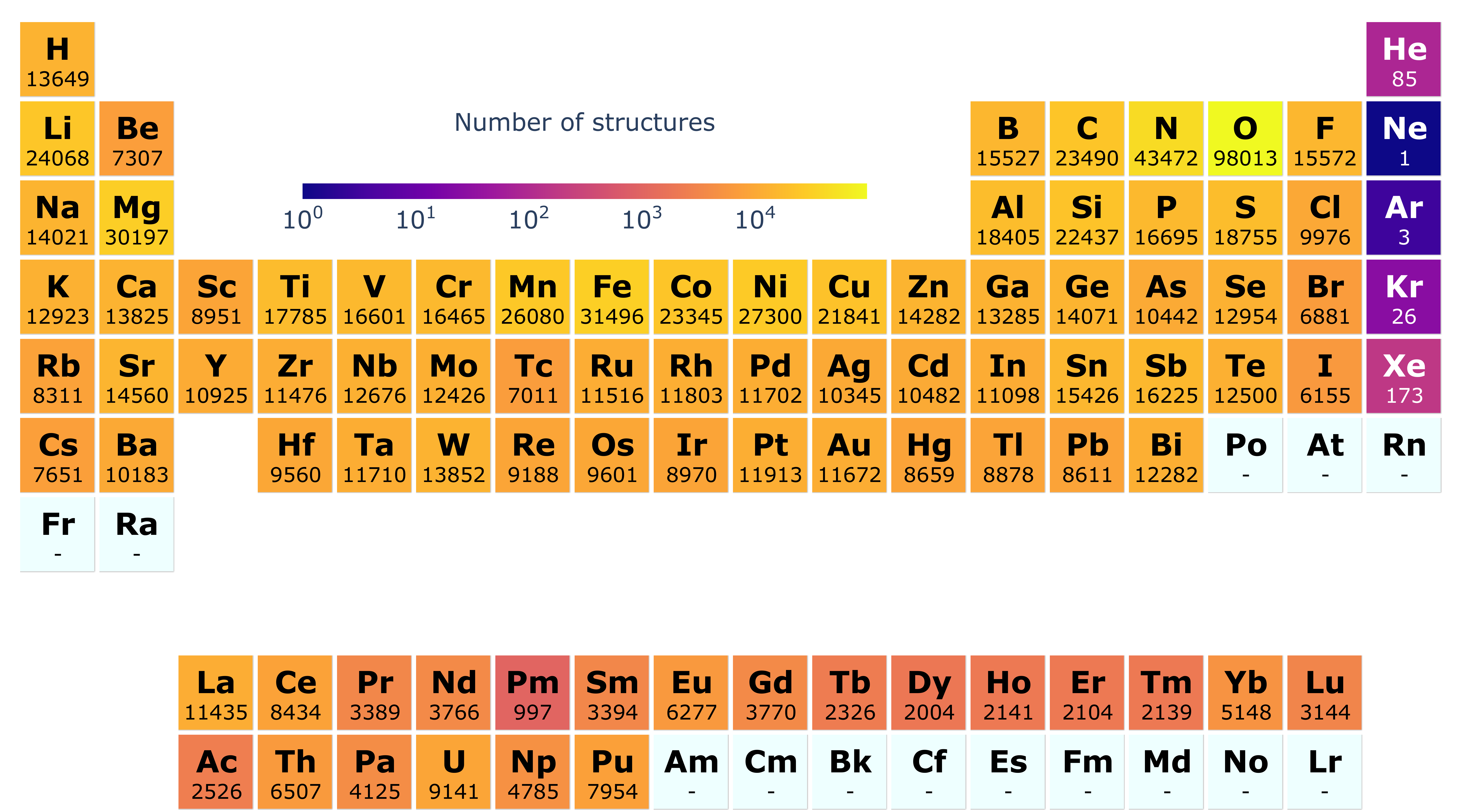}
        \caption{\label{subfig:r2scan_ele_heat}}
    \end{subfigure}
    \begin{subfigure}{0.49\columnwidth}
        \centering
        \includegraphics[width=\columnwidth]{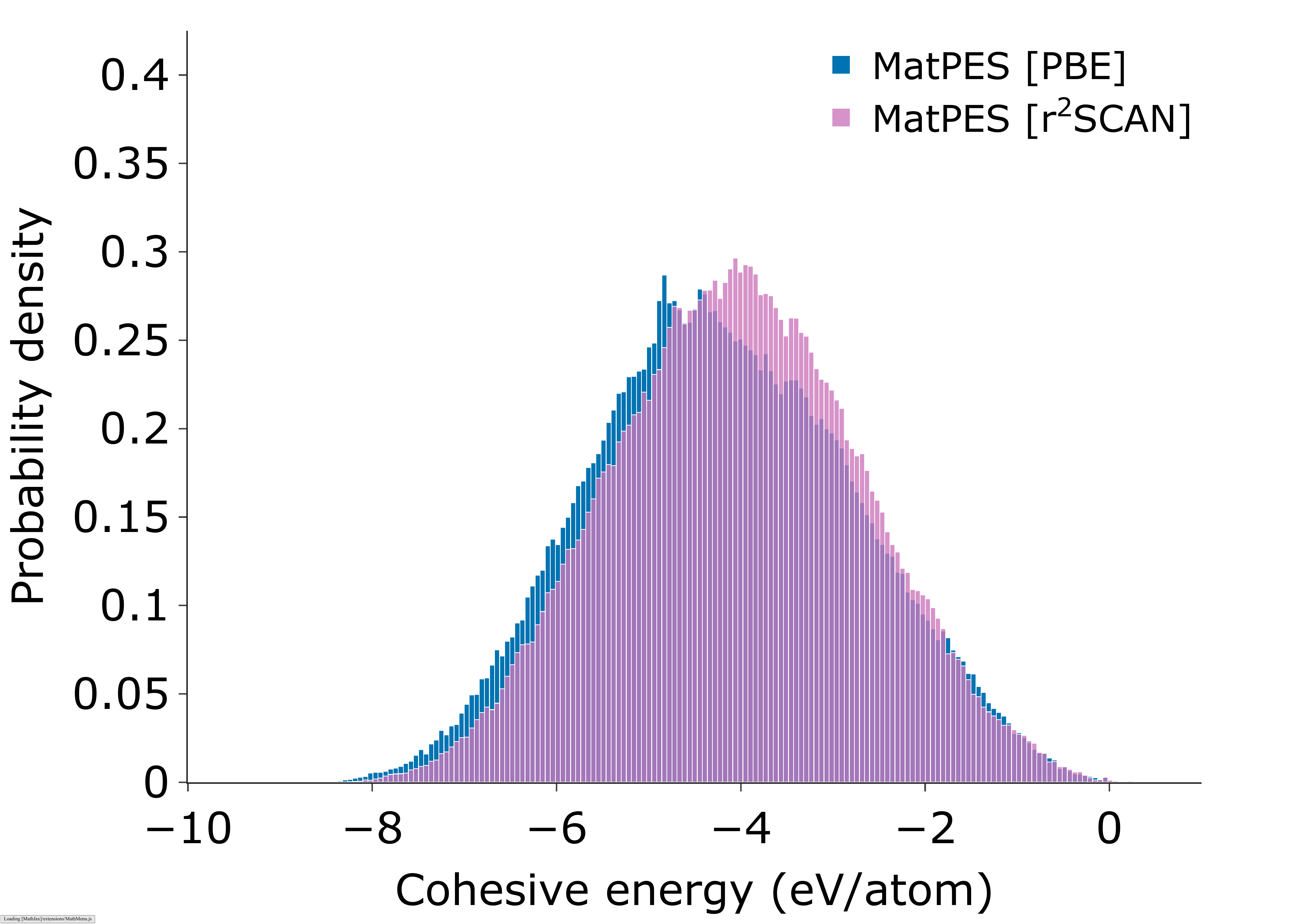}
        \caption{\label{subfig:matpes_e_coh}}
    \end{subfigure}
    \begin{subfigure}{0.49\columnwidth}
        \centering
        \includegraphics[width=\columnwidth]{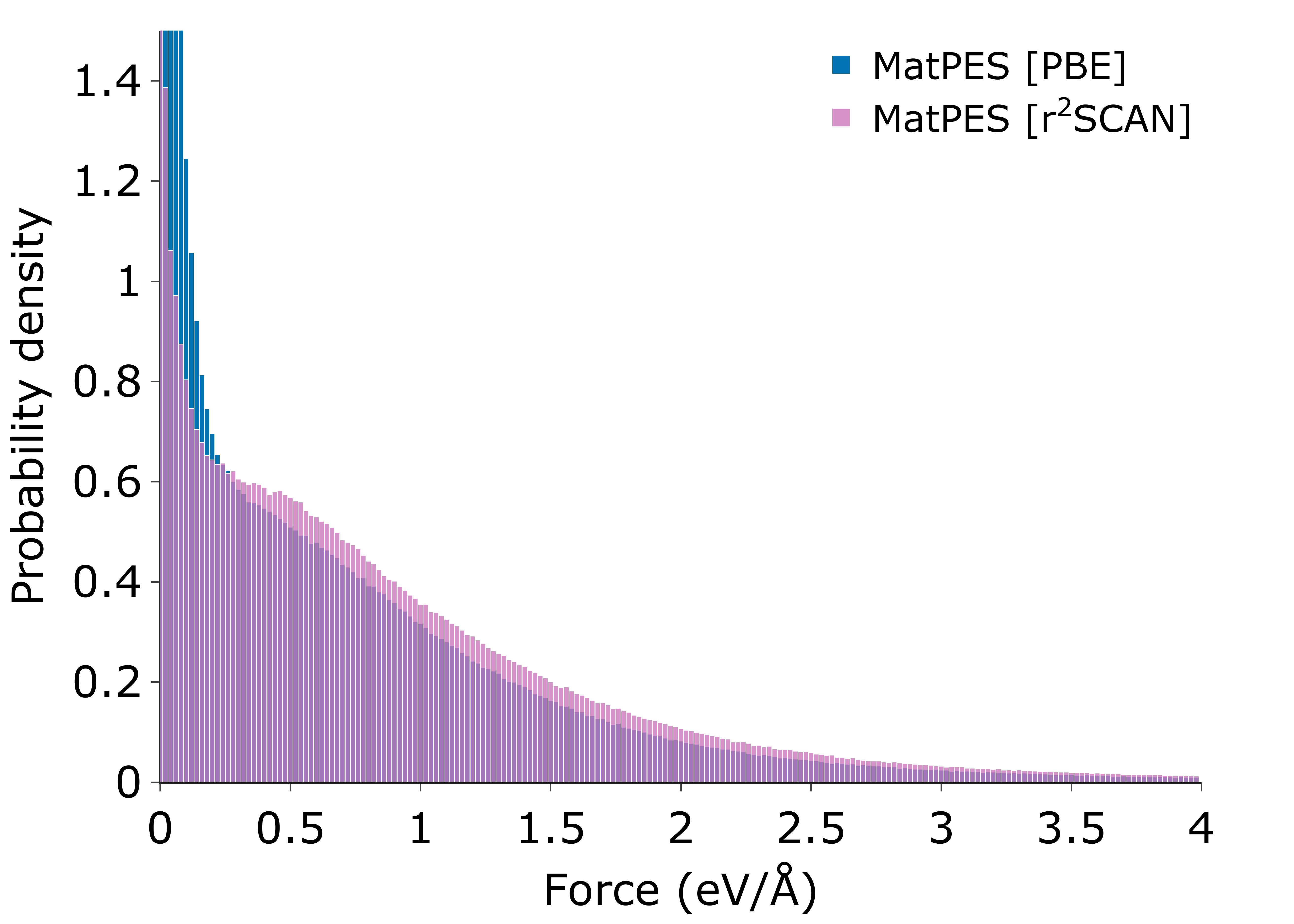}
        \caption{\label{subfig:matpes_force}}
    \end{subfigure}
    \caption{
        \textbf{Coverage of the MatPES \rrscan dataset.} \textbf{a,} Heat map of the element distribution in the MatPES \rrscan dataset. The number in each cell is the number of structures in the MatPES \rrscan dataset containing that element, plotted on a logarithmic scale. Distribution of \textbf{b,} cohesive energies (\ecoh) and \textbf{c,} interatomic force magnitudes (\absforce) in the MatPES \rrscan dataset (purple), with the PBE dataset (blue) plotted for comparison.
        The composition of the datasets are as follows: MatPES PBE: 434,712 structures (326,635 MD snapshots, 108,077 MP equilibrium structures); MatPES \rrscan: 387,897 structures (302,373 MD snapshots, 85,524 MP equilibrium structures).
    }
    \label{fig:data_comp_r2scan}
\end{figure}

\begin{figure}[htp]
     \centering
     \includegraphics[width=0.7\linewidth]{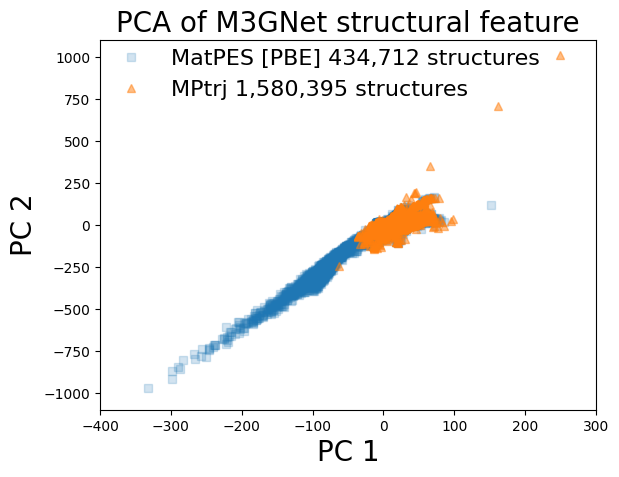}
     \caption{\label{fig:PCA_MatPES_vs_MPtrj}\textbf{Coverage of MatPES vs MPtrj datasets.} Plot of the first two principal components of the structures in the MatPES and MPtrj datasets, using the principal component analysis trained on the structural features on all MD snapshots. It is clear that the MatPES dataset covers a much range in the PC space.}
 \end{figure}

\begin{table}[h]
    \centering
    \begin{tabular}{l|c|cc|cc|cc|l} \hline
& Number & \multicolumn{2}{c}{\ecoh (eV/atom)} & \multicolumn{2}{|c}{\absforce (eV/\AA{})} & \multicolumn{2}{|c}{Pressure (GPa)} & \multicolumn{1}{|c}{Ref} \\ \hline
 & & $\mu$ &  $\sigma$ & $\mu$ &  $\sigma$ & $\mu$ &  $\sigma$ & \\ 
MPF & 185,877 & -4.273 & 2.035 & 0.527 & 4.440 & -2.374 & 37.768 & [\citenum{chen2022m3gnet}]\\ 
MPtrj & 1,580,395 & -4.325 & 1.254 & 0.327 & 1.361 & 0.281 & 6.217 & [\citenum{deng2023chgnet}]\\ 
OMat24  & 100,824,585 & -4.610 & 2.000 & 1.970 & 4.630 & 2.403 & 9.613 & [\citenum{barrosoluque2024omat}]\\ 
MatPES PBE  & 434,712  & -4.163 & 1.432 & 0.811 & 3.274 & 1.596 & 10.981 & This work\\ 
MatPES \rrscan & 387,897  & -4.008 & 1.352 & 0.935 & 2.951 & 0.893 & 11.482 & This work\\   \hline 
\end{tabular}
    \caption{
    \textbf{Number of structures, mean ($\mu$) and standard deviation ($\sigma$) of the cohesive energies \ecoh, interatomic force magnitudes \absforce and pressure for the different datasets.} The MatPES PBE dataset has mean \ecoh and \absforce that are between those of the MPF/MPtrj and OMat24 datasets.
    }
    \label{tab:dataset_stats}
\end{table}

\begin{figure}[htp]
\centering
\includegraphics[width=\columnwidth]{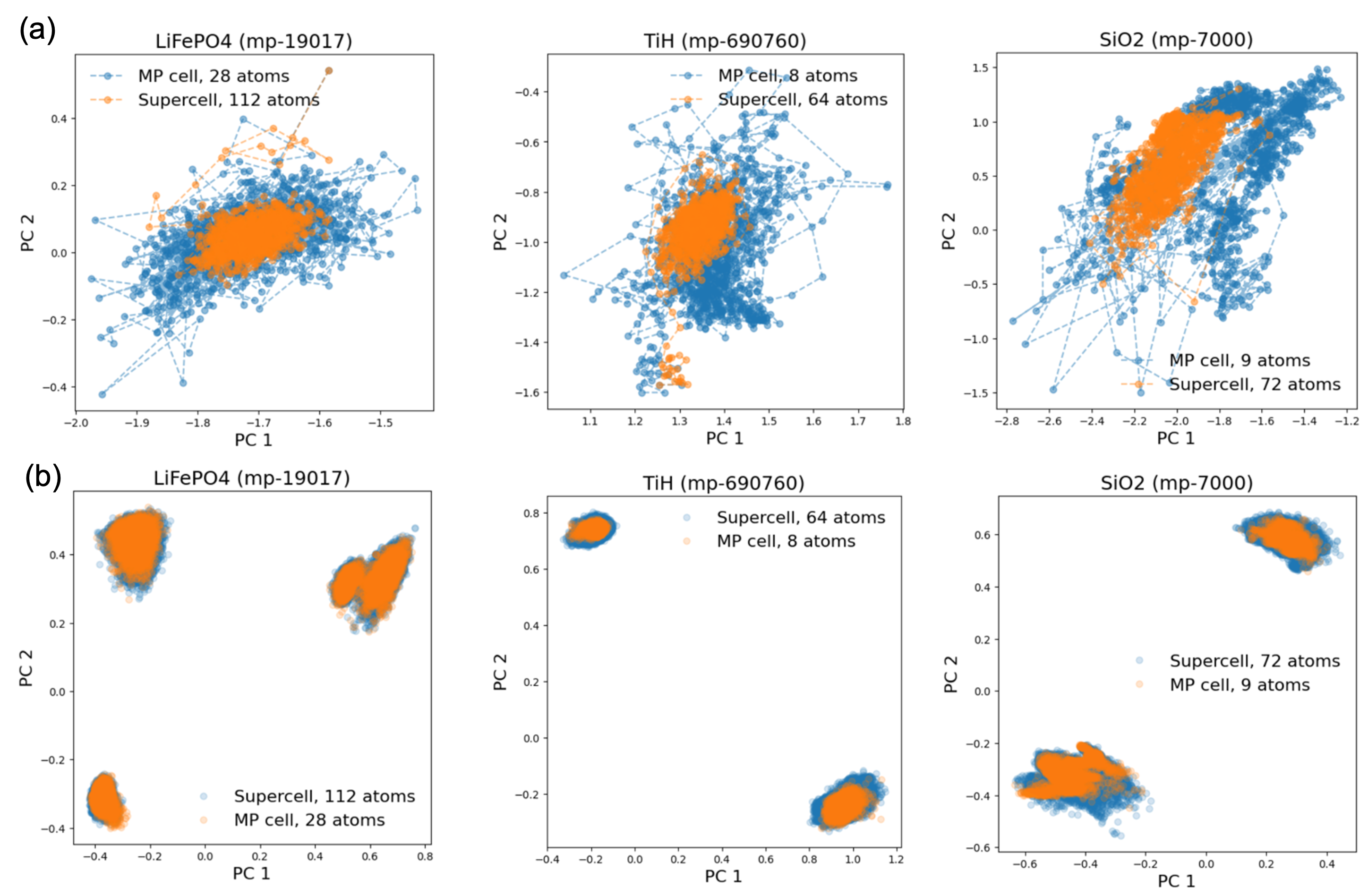}
\caption{\textbf{Coverage of structural and atomic feature space by supercells and unit cells.} Two dimensional principal component analysis (PCA) feature space of M3GNet for a, structural and b, atomic features sampled in 100 ps of $NpT$-MD at 300 K and 1 atm for three representative materials - \ce{LiFePO4} (mp-19017), \ce{TiH2} (mp-690760), and \ce{SiO2} (mp-7000). The MD supercells cover a smaller region in in structural feature space than the corresponding unit cells, due to the normalization over a larger number of atoms. However, the supercells cover a larger region in atomic feature space than the corresponding unit cells.
}
\label{fig:feature_space_supercell_unitcell}
\end{figure}

\clearpage

\begin{figure}
    \centering
    \begin{subfigure}{0.9\columnwidth}
        \centering
        \includegraphics[width=\columnwidth]{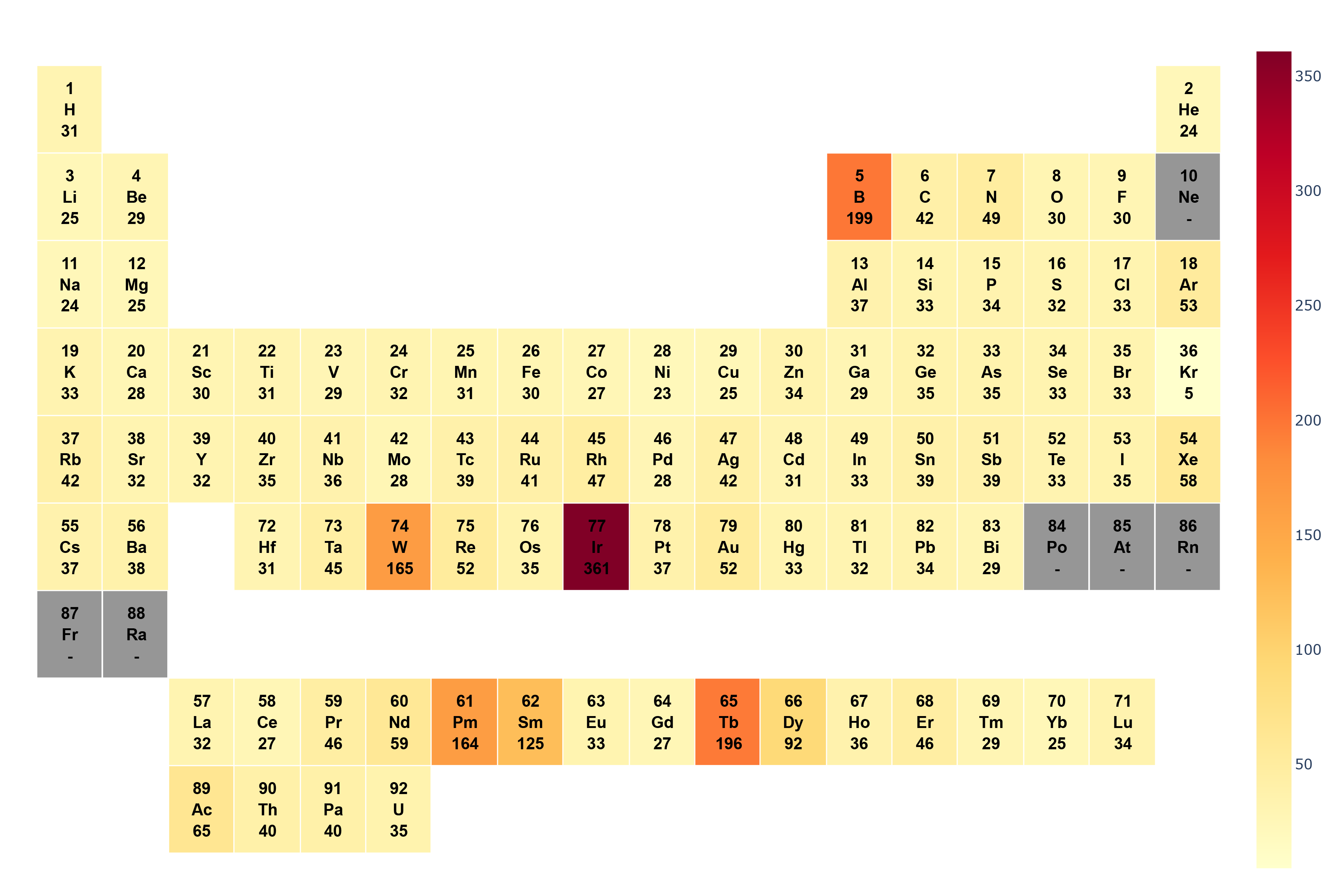}
        \caption{\label{subfig:matpes_mae_heat}}
    \end{subfigure}
        \begin{subfigure}{0.9\columnwidth}
        \centering
        \includegraphics[width=\columnwidth]{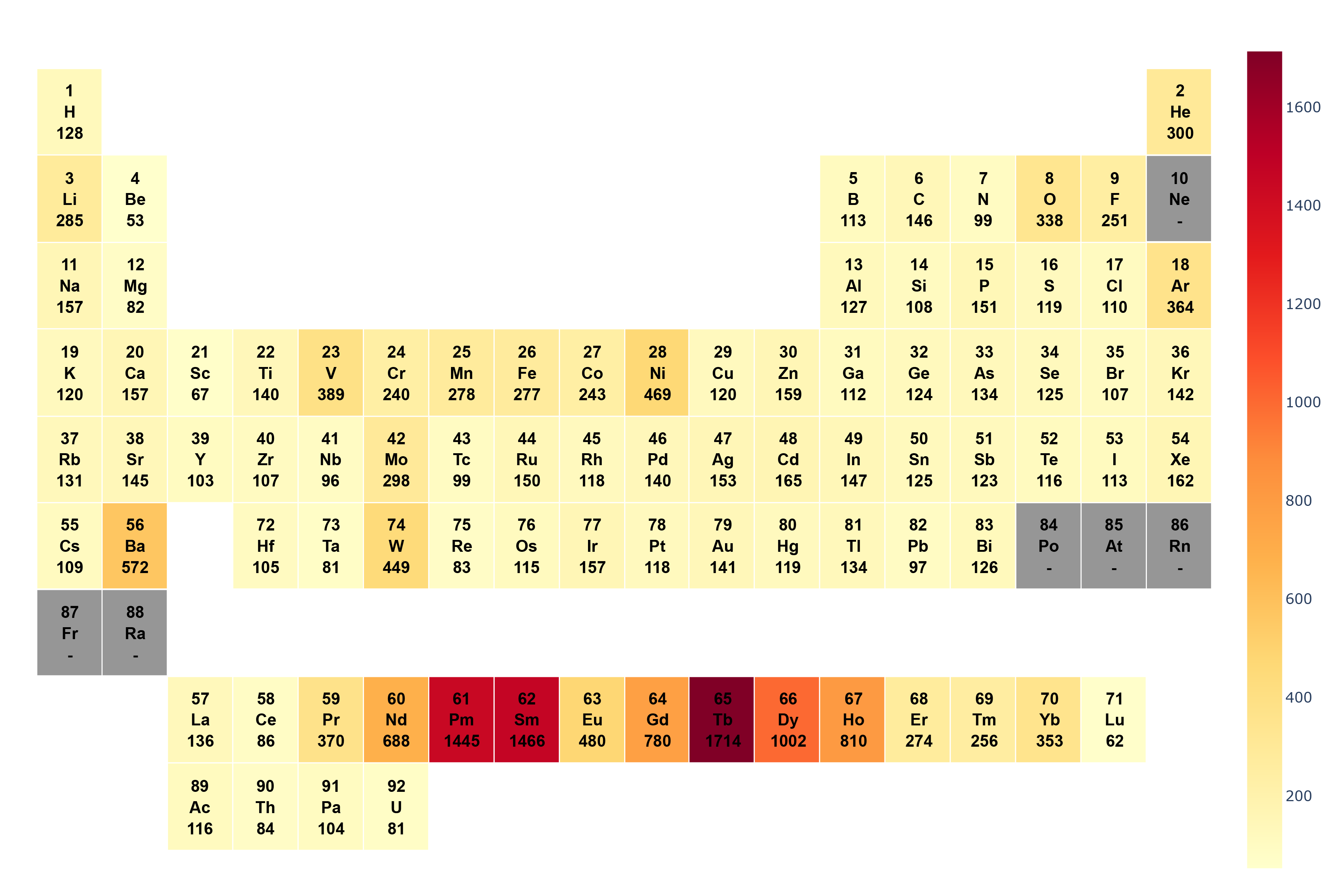}
        \caption{\label{subfig:omat24_mae_heat}}
    \end{subfigure}
    \caption{\textbf{Elemental heatmap of test mean absolute errors (MAEs) in energies.} MAEs shown are for TensorNet UMLIPs trained on \textbf{a,} MatPES PBE and \textbf{b,} OMat24 datasets.}
    \label{fig:mae-el-heatmap}
\end{figure}

\clearpage

\begin{table}[h]
    \centering
    \caption{\textbf{Overview of MatCalc-Benchmark metrics.} They can be divided into three categories: equilibrium, near-equilibrium, and molecular dynamics properties. The time per atom per time step $t_{\mathrm{step}}$ was computed using  MD simulations conducted on a single Intel Xeon Gold core for a system of 64 Si atoms under ambient conditions (300 K and 1 bar) over 50 ps with a 1 fs time step. These properties were all computed using the \texttt{MatCalc} library. \cite{matcalc}}
    {
    \begin{tabular}{lccc} 
        Task & Functional & Test Data Source & Number \\
        \hline
        \textbf{Equilibrium}\\
        \multirow{2}{*}{Structural similarity} & PBE     & WBM\cite{wangpredicting2021}& 1,000   \\
        & \rrscan & GNoME\cite{merchant2023gnome} & 1,000 \\
        \multirow{2}{*}{Formation energy per atom (\eform)} & PBE& WBM   & 1,000   \\
        & \rrscan & GNoME & 1,000 \\
        \textbf{Near-equilibrium}\\
        Bulk modulus ($K_{\mathrm{VRH}}$)                    & PBE     & MP\cite{jain2013mp}               & 3,959 \\
        Shear modulus ($G_{\mathrm{VRH}}$)                   & PBE     & MP                                & 3,959 \\
        Const. vol. heat capacity ($C_V$) & PBE     & Alexandria\cite{loewphonon2024} & 1,170 \\
        Off-equilibrium force ($|\mathbf{F_i}|$)     & PBE         & WBM high energy states\cite{deng2024soft} & 979 \\ 
        \textbf{Molecular dynamics}\\
        Median termination temp ($T_{1/2}^{\mathrm{term}}$)               & PBE \& \rrscan & This work & 172 \\
        Ionic conductivity ($\sigma$)         & PBE         & This work & 698 \\
        Time per atom per time step ($t_{\mathrm{step}}$) & PBE \& \rrscan & This work & 1 \\
    \end{tabular}
    }
    \label{tab:benchmarking_tasks}
\end{table}

\clearpage

\begin{table}[h]
    \centering
    \caption{\textbf{Computational cost of different UMLIPs.} The time per atom per time step was estimated from MD simulations conducted on a single Intel Xeon Gold core for a system of 64 Si atoms under ambient conditions (300 K and 1 bar) over 50 ps with a 1 fs time step. The total learnable parameters of the M3GNet, TensorNet, and CHGNet models used in this work are 664,000, 838,000, and 2,700,000, respectively. A clear correlation is seen between the computational cost and the number of model parameters.}
    \begin{tabular}{lc} 
        UMLIP & Time (ms/atom/step) \\
        \hline
             \textbf{MatPES PBE}\\
             M3GNet    & 1.957\\
             CHGNet    & 4.915\\
             TensorNet & 3.014\\
            \textbf{MatPES \rrscan}\\
             M3GNet    & 1.943\\
             CHGNet    & 4.875\\
             TensorNet & 3.052\\
             \textbf{MPF}\\
             M3GNet    & 2.095\\
             TensorNet & 1.803\\
             \textbf{MPtrj}\\
             CHGNet    & 4.397\\
             \textbf{OMat24}\\
             TensorNet & 3.114
            \end{tabular}
    \label{tab:speed_differences}
\end{table}

\clearpage
\begin{table}[htp]
    \centering
    \begin{tabular}{lr} \hline
    Parameter & Value \\ \hline
ALGO & Normal \\
EDIFF & 1e-05 \\
ENAUG & 1360 \\
ENCUT & 680 \\
ISMEAR & 0 \\
ISPIN & 2 \\
KSPACING & 0.22 \\
LAECHG & True \\
LASPH & True \\
LCHARG & True \\
LDAU & False \\
LDAUTYPE & 2 \\
LMAXMIX & 6 \\
LMIXTAU & True \\
LORBIT & 11 \\
LREAL & False \\
LWAVE & False \\
NELM & 200 \\
NSW & 0 \\
PREC & Accurate \\
SIGMA & 0.05 \\
\hline
\end{tabular}
    \caption{
    VASP \texttt{INCAR} settings used for MatPES.
    For PBE \cite{perdew1996pbe} calculations, the \texttt{GGA} tag was set to ``PE'', and for \rrscan \cite{furness2020r2scan} calculations, the \texttt{GGA} tag was not set, and the \texttt{METAGGA} tag was set to ``R2SCAN''.
    }
    \label{tab:matpes_incar}
\end{table}

\begin{landscape}

\begin{longtable}[!h]{lrr}
  \caption{
  \textbf{VASP pseudopotentials (\texttt{POTCAR}) used for MatPES.}
  The second column indicates the ``TITEL'' keyword of the POTCAR, and the third column indicates the SHA256 hash that is included in the POTCAR file.
  MatPES uses the ``PBE 64'' PAW pseudopotential library.
  \label{tab:matpes_potcar}
  }

\\ \hline 
El. & TITEL & SHA \\ \hline  \endfirsthead
\\ \hline 
El. & TITEL & SHA \\ \hline  \endhead    
\hline \endlastfoot
Ac & PAW\_PBE Ac 06Sep2000 & ef0c2b83cf569bea36d28252deb147ae18f4f417709a0a90a00fa0751e60408a \\
Ag & PAW\_PBE Ag 02Apr2005 & 6550cfa3543261e132c169e1b98a529204624d72be14e2da2e0242bcb74a174d \\
Al & PAW\_PBE Al 04Jan2001 & 17880443556af62b473fe41b62a467bd001ad55d2cabe504a3f22e34d4e9db96 \\
Am & PAW\_PBE Am 08May2007 & ed4d25cb37bf36722bf7da53aff49a5f32f6e6cb278d4cb693f29442583ade28 \\
Ar & PAW\_PBE Ar 07Sep2000 & 94ccc759e5215f718956dfe5fb43b4f2f2ae700f97accfbf846e1c2491cbae56 \\
As & PAW\_PBE As 22Sep2009 & bc8fb55b00baa90d383a523722e1771deb40ea3f17a5ce25913641995975acad \\
At & PAW\_PBE At 21May2007 & 324a31b576a03b2f68d883a08822bc631d1aee72d359804fa17db26f45fb52bb \\
Au & PAW\_PBE Au 04Oct2007 & d0044ae04e2bdce24051b198fc5c053d722a5bc6fe3c3b100514a13fc5d2db88 \\
B & PAW\_PBE B 06Sep2000 & a32ced30f5ae56fd4d10b4325ff17eb3e4e38ee0f4288bc219fb012fddfa6e97 \\
Ba & PAW\_PBE Ba\_sv\_GW 30Nov2021 & 729cfb57c5620ba7c0f6e42203de8340b45f71b59c373a5821ca22f69225df8c \\
Be & PAW\_PBE Be\_sv 06Sep2000 & 95d73059eaef0de9a2d42225277cca956d5b2c38504d1b500b1a4c08b9931b7a \\
Bi & PAW\_PBE Bi 08Apr2002 & d6b6753ed5db3f0e277fb15e6dbb6699c4bc829850a481068e9e7236faeca489 \\
Br & PAW\_PBE Br 06Sep2000 & 96a73d2954943bbee26f4990d676cc6c3bf44b8dd2c75cd4b3b825d8403f0103 \\
C & PAW\_PBE C 08Apr2002 & 253f7b50bb8d59471dbedb8285d89021f4a42ed1a2c5d38a03a736e69125dd95 \\
Ca & PAW\_PBE Ca\_sv 06Sep2000 & a47365830e737f14e0e6c5cf1ed81b94e081eecf0a33df105380881bc9da05d5 \\
Cd & PAW\_PBE Cd 06Sep2000 & 8b7ca71966beae5276c8bb910adb3ecc013a3354c27473914c94cd54c83be4f7 \\
Ce & PAW\_PBE Ce 23Dec2003 & 00bd3101dba980d69718c826e9aff48526de61f93249f80fb0d1cde9afca69b7 \\
Cf & PAW\_PBE Cf 17Oct2013 & 9d2a2d228fc4747daad0b17273e4831741ec188f58739a971ab0474c5ff36db3 \\
Cl & PAW\_PBE Cl 06Sep2000 & 9f1b6e6ed4247ac726a768b330d26d7667cc3c95a39ff49c1d064c2d9dcce931 \\
Cm & PAW\_PBE Cm 17Jan2011 & 10f7147aec31bdfcd023378638ce6d89b1eef3e653eb06887ddc6b577c1a20c8 \\
Co & PAW\_PBE Co 02Aug2007 & 0e690c60710354995174544f52f9f2c30879afabeffe3b3fbd4001cb294e56d4 \\
Cr & PAW\_PBE Cr\_pv 02Aug2007 & c9a7df34d3cdbacf1090e328ef39c0b420964e11c23f63548d1ac9dd218cdba0 \\
Cs & PAW\_PBE Cs\_sv 25Jan2019 & 0e16bce67778f8d3e6a1e8d49098acae95c0f8d441dd09aee70633bcf454619d \\
Cu & PAW\_PBE Cu\_pv 06Sep2000 & cb7b504e2ea725fe1f25c85a9ac77d4012ce94cd394135b722c4e25ec297f1cb \\
Dy & PAW\_PBE Dy\_h 30May2022 & 22476d747c0ad3010bbc2b9b82ce7b879d05e04135e746a70820417bee947c38 \\
Er & PAW\_PBE Er\_h 29Jun2022 & 2b80424db8faabd5254b489548b03ac32ffe345d20d95add1e30ba732b45f18a \\
Eu & PAW\_PBE Eu 25May2022 & 60d5a46ff9a0a7f4ab06309e7661032aad63da4a27607063648b9a9b69393f0f \\
F & PAW\_PBE F 08Apr2002 & 53c630871ac675939349fc2b976745ee17808dedc90b5bdde026bc81f0faf456 \\
Fe & PAW\_PBE Fe\_pv 02Aug2007 & 5d22e414b1f82158bf2c7ecb8b97b28fd0923e48cadc1c3bf74d524558f5dd32 \\
Fr & PAW\_PBE Fr\_sv 29May2007 & 23d9c34aa2eb6adab1bca1262c635f9251f4b1fe31f8522b3106ee5dd2f6057e \\
Ga & PAW\_PBE Ga\_d 06Jul2010 & a60ddf36e14f00ea098d9e3914b3745f1b7105e68786148f49ae384e44b4226c \\
Gd & PAW\_PBE Gd 25May2022 & b942d524b9340ee44669b7c5645128971170deeaff9d9af5b36d2c15a338fd71 \\
Ge & PAW\_PBE Ge\_d 03Jul2007 & 944b26c40d2d7c4f4eb1ff3a2e8dbfdd7129146276cd341854bfd5c0e57780ed \\
H & PAW\_PBE H 15Jun2001 & 030f79b5d3ab3cf0e668861823c8fb652ff669f3e15e46930bd03bfd63a607b6 \\
He & PAW\_PBE He 05Jan2001 & 767818bb8a862153b2ebc238b4abd4bed99a882bbb8e6a4800cddfa4f1a760c3 \\
Hf & PAW\_PBE Hf\_pv 06Sep2000 & 326372999ee61732e8151b0274b0330bf639aff42dca34b327393d3d5ff5d3de \\
Hg & PAW\_PBE Hg 06Sep2000 & b58054e5facb8a6c456f8fea289fc655b681c0fa06131ba074282c377c596e89 \\
Ho & PAW\_PBE Ho\_h 29Jun2022 & f964032ac636ca1bab774bb902286f5c66af571df067f67ca2e3f0edddbd1c51 \\
I & PAW\_PBE I 08Apr2002 & e40f3f59b681c1fc3e3091736183c4589116986673d2b852a5012aec72758799 \\
In & PAW\_PBE In\_d 06Sep2000 & bef4eefb233e1f458c7ef2e09e39e7114bb5a6a38c7ac356915527764429c513 \\
Ir & PAW\_PBE Ir 06Sep2000 & 7c6af8d4d487b237782eb51e6f62977b27a1da78c3c96acd1cebefd6c308f120 \\
K & PAW\_PBE K\_sv 06Sep2000 & bf8373ef592e31d27efa2dcc68371be6d6a25ce4db6c2ffaf9e92c44050ba21a \\
Kr & PAW\_PBE Kr 07Sep2000 & 6b89d4ab453c74a018c642cd8d7c9a2bec7c33c7e53fbff7492fc5bbd2c9051e \\
La & PAW\_PBE La 06Sep2000 & b7aad99517e50aeb53363b20a29bb0dfa896544080812fb23322f071df953199 \\
Li & PAW\_PBE Li\_sv 10Sep2004 & 7e51fe1804c037e1dccc81a9c376d94d693a7559600c847f4b41960edb8ab895 \\
Lu & PAW\_PBE Lu\_3 06Sep2000 & fe1f8b446106b829a34406137eadf9babd792fb03d1986447d054dbaf8059d6c \\
Mg & PAW\_PBE Mg\_pv 13Apr2007 & f474ac0dd33840b9ae76c01d57fea79fdf77cdfbb07d5ade72c65f83c709b62b \\
Mn & PAW\_PBE Mn\_pv 02Aug2007 & c79df11ab18a0e3347296df6ded47bb5f18b2e4fbd621d0b5280d2eae24c30a0 \\
Mo & PAW\_PBE Mo\_pv 04Feb2005 & 6ae1433eb25a8c9ce9b558610a2d5a3c8775e861c8272ad13d852ff7f9633ae0 \\
N & PAW\_PBE N 08Apr2002 & e053789ff3a61a86a1b75d8a110fcc91f86041011e7b0817c7c99e4e8a6349d7 \\
Na & PAW\_PBE Na\_pv 19Sep2006 & 6a2f546d9e11350984debacf3dc457d8cffc0868e817d445faca461816a32b94 \\
Nb & PAW\_PBE Nb\_pv 08Apr2002 & bac2b60850b34f8515cf56f9feae68b678e347865051052a0e1f5d2c6a691c0a \\
Nd & PAW\_PBE Nd\_h 01Jun2022 & 7bd4bb7cee51b5dc2e0f9cbae2e86e657034401e01318a3a9244a0deed38a434 \\
Ne & PAW\_PBE Ne 05Jan2001 & 7551ec1d38f079f813f98269ad695dc650bd9c34a5dedf814d6a76328defc8c0 \\
Ni & PAW\_PBE Ni\_pv 06Sep2000 & 368cd815a19284a5fda64519f43fef792dad8376deb9e2da41aae78b997dc50e \\
Np & PAW\_PBE Np 06Sep2000 & ccc3f89c89c668b33a1cedb3a141ff87191bee6923db3e8755910d8c985d3af2 \\
O & PAW\_PBE O 08Apr2002 & 818f92134a0a090dccd8ba1447fa70422a3b330e708bb4f08108d8ae51209ddf \\
Os & PAW\_PBE Os\_pv 20Jan2003 & df2e3ef880fc2502babe687ce20ec2e4b82fab5eb8dd3b5c0a23d3b7861e7a8b \\
P & PAW\_PBE P 06Sep2000 & df60c54a93efe35c9e85ae94c010c75e9f6960a95d5595d7cdba4096d109af88 \\
Pa & PAW\_PBE Pa 07Sep2000 & 04e32654b760de29c7f02015f2c4b9b92aeb120699210996a29249ba80bfa323 \\
Pb & PAW\_PBE Pb\_d 06Sep2000 & fb885b08f0fba73a15f6a6b0e39bd200c13ec96ef97cf3ecd8e63bade5137abd \\
Pd & PAW\_PBE Pd 04Jan2005 & dd6f6f02930356371984e2b707ff5e456046761dc76b07c1eba30abd15eb3c39 \\
Pm & PAW\_PBE Pm\_h 01Jun2022 & e20fde408f4bc5ec1ab6d6cec4ca07cb81fc0f789787f0fe8d07d98759b2aec6 \\
Po & PAW\_PBE Po\_d 25May2007 & 589bb7cf7db41d81724fa452df1c3d870736441cf2fe2863a84270f4db2d0320 \\
Pr & PAW\_PBE Pr\_h 01Jun2022 & 0607a17d0060e989022cd48f8f145dff5f704ed14e44fe020789997c290d1ea3 \\
Pt & PAW\_PBE Pt 04Feb2005 & 3ed90460adef76debcff1cfb73ee1349b515b1a3db03439735b44de3a8db7dc8 \\
Pu & PAW\_PBE Pu 06Sep2000 & aea5004a3542f2b7cbf52448d5c3d9b9e438d6248055eeb36c781f43d2b3cf3c \\
Ra & PAW\_PBE Ra\_sv 29May2007 & c89b71e1d92b290ba352aa2322e5ac7f7cc2f71bf48f725710403fd01d027668 \\
Rb & PAW\_PBE Rb\_sv 06Sep2000 & 06c38fe1ec7709d96d3959f5f2c66fec330a93b52b042a8fb2dfe6a3a3fb06c1 \\
Re & PAW\_PBE Re\_pv 06Sep2000 & 9136fabd3a1e35b9fb4b1f356ef358c338d3d4e46f3e96ab15e45974af27d0c0 \\
Rh & PAW\_PBE Rh\_pv 25Jan2005 & 25b11608b0d10a93adcc8487bc062d7eb8aeda443ca523c31dcc1b5e75119ef1 \\
Rn & PAW\_PBE Rn 12Aug2016 & 1082a7f1e478858715023a6a817d2c8d5512517dec9bddb12f27eca6306ec2cc \\
Ru & PAW\_PBE Ru\_pv 28Jan2005 & 539bee49bb4e63d8933d8244c4f66bec34c4e5cbe2759ec5d366ad1017752494 \\
S & PAW\_PBE S 06Sep2000 & 0fc7481fb0695f01bdc6462160264c5c84044ae9ec85a907d398b887a2bc3132 \\
Sb & PAW\_PBE Sb 06Sep2000 & 8a1325a3afd8ca988779475cff04188eb880a428d1acb53275fe456fa3b784fc \\
Sc & PAW\_PBE Sc\_sv 07Sep2000 & 9aad5a0618293b7e22b0823afd8bf80ef5c5e525eb8618ba91672918e18a06fa \\
Se & PAW\_PBE Se 06Sep2000 & eabb916e6b4c819dc065ce039373bc328651da483898f08fb9e49c498452bf12 \\
Si & PAW\_PBE Si 05Jan2001 & 79d9987ad8750f624c4d6acb2a16d13abf6a777132adc04dc6c8399be72b42bb \\
Sm & PAW\_PBE Sm\_h 30May2022 & 2f62dec1b7198d20317f663b75f15c0d34872d9e465e7140f969bed2fb3252a3 \\
Sn & PAW\_PBE Sn\_d 06Sep2000 & 385b269c1887fa92a2bd1b595c4ea1490c7eb92f9f3d87257729a4a813cde741 \\
Sr & PAW\_PBE Sr\_sv 07Sep2000 & a8389d3481648516ee67182ee030b298d280b37e29cbbf2d8595f040d757c710 \\
Ta & PAW\_PBE Ta\_pv 07Sep2000 & 00d0a14a36c127416459414afb8617d9754996fffa60a2465888e660b20a452f \\
Tb & PAW\_PBE Tb\_h 30May2022 & 1b65b31de2c27578bdfaaa2be6c07419aa8a7e4c6a2be444d54e32093864b0c5 \\
Tc & PAW\_PBE Tc\_pv 04Feb2005 & 55d1e894bb6e22d412434369d62e29007ee0956d7c4b4ca346dcdb73d72337c9 \\
Te & PAW\_PBE Te 08Apr2002 & e13b0861f25acb1fe6cdc17458a3647985498b67a429eb77ddc1eac05a1775b1 \\
Th & PAW\_PBE Th 07Sep2000 & d41e9f824f712d7814b19b6bc43381cc9a373f2ab99e53c63e55d3471ac6bd23 \\
Ti & PAW\_PBE Ti\_pv 07Sep2000 & f757a1b2c6d082f4c628fa3d987464a8763bf92e53844ac0500b0e2ddc9ce5c0 \\
Tl & PAW\_PBE Tl\_d 06Sep2000 & 114f54bd8cac727af4c1a2dcc375cc6f82ae0f808e1efd36f54552a740d0afeb \\
Tm & PAW\_PBE Tm\_h 29Jun2022 & 14c2249a7b2afb252acd0daad4b9c552cb4fdb549ec5080ecaac209e82a28e2c \\
U & PAW\_PBE U 06Sep2000 & 83ff4a3ef579a1d3def1e25ae1b036f9320912559a2f473c33031cce813da2e2 \\
V & PAW\_PBE V\_pv 07Sep2000 & 1175d3c8cb4ffd1d150a520fa74fae5c3ec72c0849b5615a94a5372dd2cf07f7 \\
W & PAW\_PBE W\_sv 04Sep2015 & 931c2d770f65867ef30f3db3421922900fc0890ebac5c3e12f63b6a2064023d7 \\
Xe & PAW\_PBE Xe\_GW 08Jan2009 & d550ac1633c6a0fdbf4cc3b9e9481b749e6bbfc02e3e0a4eed1c4f3493688506 \\
Y & PAW\_PBE Y\_sv 25May2007 & 5e7f7496c6fa99024fec326bcd60dcee8ff7e886f9613d31da35c685a538a036 \\
Yb & PAW\_PBE Yb\_h 29Jun2022 & cfae7690412fd84d754fbf73d8a1e0d0ef26063421fabf4cb7e6c45490da7400 \\
Zn & PAW\_PBE Zn 06Sep2000 & 501fddbd8274dd8e9725d4f3de27861d7f1e553a7e18441bc79fed8f346e7f23 \\
Zr & PAW\_PBE Zr\_sv 04Jan2005 & 25aed69cb10325f9d37c5c68912b61a17387d1f8e4f1d804860ffa10c8a4bf76 \\
\end{longtable}
\end{landscape}

\begin{figure}
     \centering
    \begin{subfigure}{0.49\columnwidth}
        \centering
        \includegraphics[width=\columnwidth]{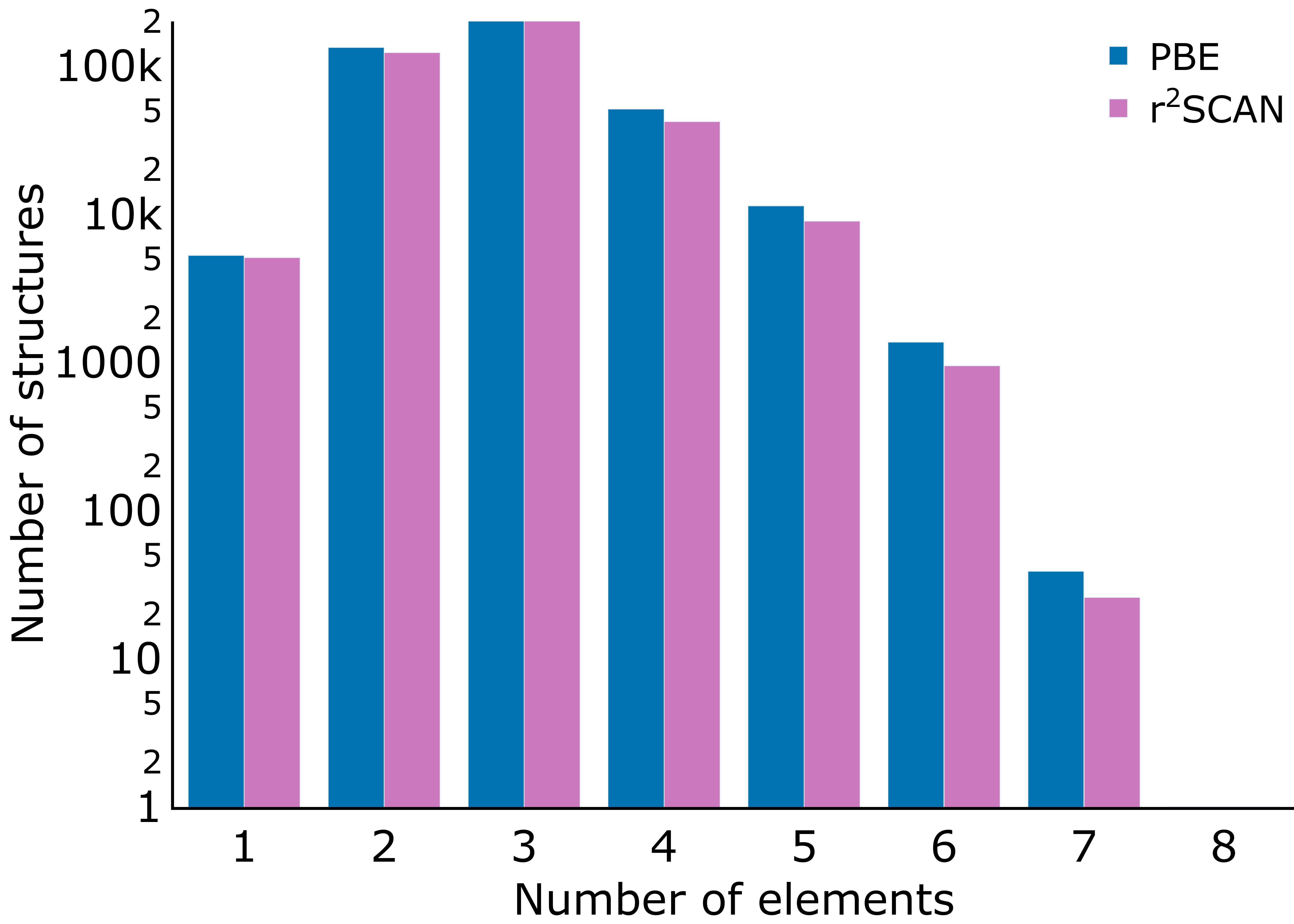}
        \caption{\label{subfig:matpes_num_ele}}
    \end{subfigure}
        \begin{subfigure}{0.49\columnwidth}
        \centering
        \includegraphics[width=\columnwidth]{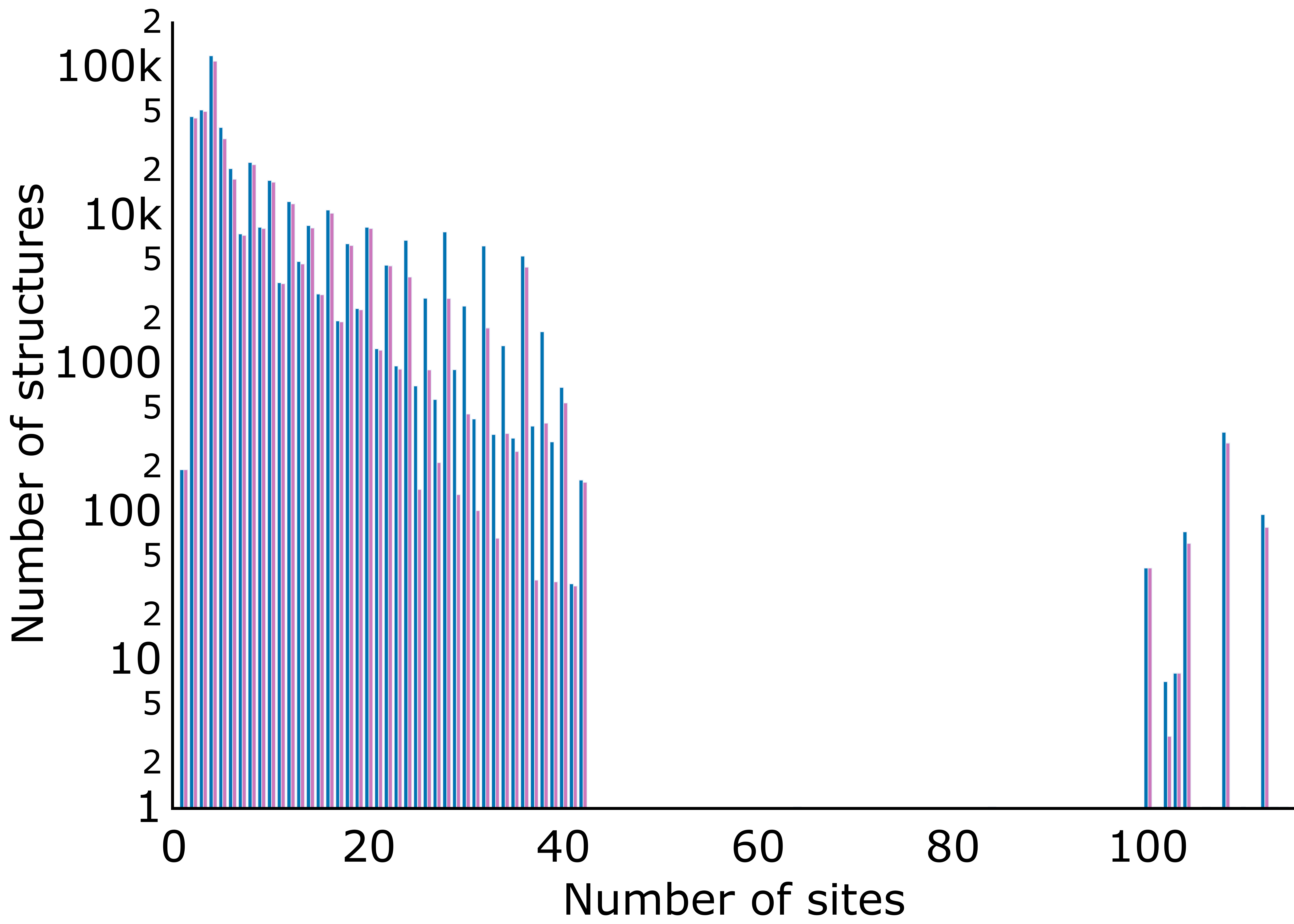}
        \caption{\label{subfig:omat24_num_sites}}
    \end{subfigure}
    \caption{\textbf{Distribution of the number of elements (a) and number of sites (b) for the MatPES structures.}
    The number of structures for a given quantity are plotted on a logarithmic scale.
    The blue (purple) bars show the structure counts for the PBE (\rrscan) subsets of MatPES.
    }
    \label{fig:ele_site_dist}
\end{figure}

\end{document}